\RequirePackage{fixltx2e}
\documentclass[a4paper]{isp}
\usepackage[american]{babel}
\usepackage{graphicx}
\usepackage{amsmath,amsthm,mathtools}
\usepackage{url}
\usepackage[capitalise,nameinlink]{cleveref}
\usepackage{xcolor}
\def\aj{AJ  }%% The Astronomical Journal
\def\apj{ApJ\,  }%% Astrophysical Journal
\def\apjl{ApJ Letters,  }%% Astrophysical Journal, Letters
\def\sn1987a{SN \,1987A\,}
\def\s1006{SN \,1006\,}
\def\degreezan{^{\,\circ}}
\begin{document}
\pdfgentounicode=1
\title
{
Energy Conservation in the thin  layer approximation:
II. The asymmetric classic case for supernovae remnants 
}
\author{Lorenzo Zaninetti}
\institute{
Physics Department,
 via P.Giuria 1, I-10125 Turin,Italy \\
 \email{zaninetti@ph.unito.it}
}

\maketitle

\begin {abstract}
Here we extend  the conservation of energy in the framework
of the thin layer approximation
to the asymmetrical case.
Four types of interstellar medium are analysed,
in which the density follows 
an inverse square profile, a power law profile, an exponential profile
and a toroidal profile.
An analytical solution for the radius as a function of time 
and the polar angle in spherical coordinates is derived
in the case  of the inverse square profile.
The analytical
and numerical results are applied to two
supernova remnants:
SN~1987A 
and  SN~1006.
The back reaction due to the radiative losses  
is evaluated in the case of 
the inverse square profile for the surrounding medium.
Two models for the image formation are presented, 
which explain  the triple ring visible in 
SN~1987A 
and the jet feature of  SN~1006.
\end{abstract}
{
\bf{Keywords:}
}
supernovae: general
ISM       : supernova remnants
supernovae: individual (SN 1987A)
supernovae: individual (SN 1006)

\section{Introduction}

The asymmetries observed   in supernovae (SNs) and supernova remnants (SNRs) 
have been 
recently  investigated   by different approaches:
on analysing a type Ia SN that exploded around CE 1900
\cite{Borkowski2017}
reported that it is 
asymmetric in the  radio region but 
shows a bipolar  morphology in the X region.
The details of the deceleration of that SN 
in different directions
were analysed by 
\cite{Reynolds2016}. 
The symmetry of many SNRs in 6-cm and 20-cm 
Very Large Array images were measured by 
\cite{Stafford2017}.
 A  three dimensional smoothed particle
hydrodynamic simulation was  used by
\cite{Collier2017}
in order to explain the observed asymmetries in SNs.
\cite{Basurto2019} implemented a
 three-dimensional MHD numerical code
in which the
initial mass distribution of the SN  is asymmetric.
The presence of an asymmetric circumstellar medium (CSM) 
responsible for the lack of narrow lines within the first two days of
the explosion for the Type II-P supernova SN 2017gmr
was suggested by 
\cite{Andrews2019}.
The differences  in the Si II line widths in type Ia supernovae
with an asymmetric CSM
were explained by 
\cite{Livneh2019}.
An 
asymmetric explosion as seen from different
viewing angles 
was suggested  by 
\cite{Strasburger2020}
in order to explain  some anomalies
in the photospheric
velocities. 
The model of energy conservation in the thin 
layer approximation for the expansion  
of SNRs was developed in \cite{Zaninetti2020a} in
the spherical case. Here we analyse the 
asymmetrical or non-spherical case.
In order to accomplish this,
Section \ref{sec_energy_motion} 
derives  the basic differential equations
which regulate the equation of motion for SNRs 
for four types of medium.
Section \ref{sec_energy_applications}
applies those analytical and numerical results
to \sn1987a and \s1006
and Section \ref{sec_energy_image}  analyses
two models of the image formation for 
the two SNRs analysed.

\section{Energy conservation}

\label{sec_energy_motion}
A point in  Cartesian coordinates is characterized by
$x,y$ and $z$,  
and  the position of the origin
is the centre of the explosive phenomena.
The same point in spherical coordinates is characterized by
the radial distance $r \in[0,\infty]$,
the polar angle     $\theta \in [0,\pi]$,
and the azimuthal angle $\varphi \in [0,2\pi]$.
In the following, the profiles of the density considered 
are independent  of the azimuthal angle
and are functions of the distance $z$ from the  $z=0$ plane.
In   spherical  coordinates, the goal  is 
to derive  the  instantaneous radius  of expansion, $r$, as 
a function of the polar angle. 
The following approaches 
will model the CSM  around 
the point of explosion of the SN 
which  is characterized 
by a plane, $z=0$, 
which   produces  an up and down symmetry in the polar angle,
$r(\theta)=r(\pi -\theta)$,
and by 
a   polar axis, $x=0,y=0$, 
which produces a right and left symmetry
for the azimuthal angle
$r(\varphi)=r(\varphi+\pi)$.
The basic symmetries 
are now outlined
\begin{subequations}
\begin{align}
r(\theta)=r(\pi -\theta) 
\label{symmetries1}
\\  
r(\varphi)=r(\varphi+\pi)
\label{symmetries2}
\quad.
\end{align}
\end{subequations}
In other words, the numerical  simulations will be 
developed in the first quadrant ($\theta \in [0,\frac{\pi}{2}]$,
$\varphi=0$ and then applied  to the other three quadrants.

The conservation of kinetic energy in
spherical coordinates
along  the  solid angle  $\Delta \Omega$
in the framework of the thin
layer approximation  is  
\begin{equation}
\frac{1}{2} M_0(r_0;\theta) \,v_0^2 = \frac{1}{2}M(r;\theta) \,v^2 
\quad,
\label{eqnvelocity}
\end{equation}
where $M_0(r_0;\theta)$ and $M(r;\theta)$ 
are the swept masses at $r_0$ and $r$,
while $v_0$ and $v$ are the velocities of the thin layer 
at $r_0$ and $r$.
The above conservation law when written as a differential
equation is  
\begin{equation}
\frac{1}{2}\,M \left( r \right)  \left( {\frac {\rm d}{{\rm d}t}}r \left( t
 \right)  \right) ^{2}-\frac{1}{2}\,{\it M_0}\,{{\it v_0}}^{2}=0
\quad.
\end{equation}
The velocity as a function of the radius is 
\begin{equation}
v(r;r_0,v_0,\theta) =
{\frac {{r_{{0}}}^{3/2}v_{{0}}}{{r}^{3/2}}}
\quad.
\end{equation}
In the following, 
$r_0$ is the radius after which the density 
starts to decrease,
$r$ is the radius of expansion in spherical coordinates,
$\rho_c$ is the density at $r=0$, 
$z =r \,\cos(\theta)$ is the Cartesian coordinate  along the $Z$ axis
and $\cos(\theta)$ is the polar angle in spherical coordinates.
The main astrophysical assumption adopted here is that 
the density of the CSM
decreases 
as a function of the distance from the centre,
due to the 
previous stellar winds.
At the time of writing, there are no
astronomical observations which outline such an effect, 
and therefore 
we test different profiles for the density and
evaluate whether they are compatible with the 
observed sections of SNRs.

\subsection{An inverse square profile of the density}

\label{section_inversesquare}

We  assume that the medium 
around the SN
scales with the axial piecewise dependence
\begin{equation}
 \rho (r;r_0)  = \{ \begin{array}{ll}
            \rho_c                      & \mbox {if $r \leq r_0 $ } \\
            \rho_c (\frac{r_0}{z})^2    & \mbox {if $r >     r_0 $}
            \end{array}
\label{piecewiseinverse}
\quad.
\end{equation}
The mass $M_0$
swept in the interval [0,$r_0$]
is
\begin{equation}
M_0(\rho_c,r_0) =
\frac{4}{3}\,\rho_{{c}}\pi \,{r_{{0}}}^{3}
\quad.
\nonumber
\end{equation}
The total mass $ M(r;r_0,\rho_c)$
swept in the interval [0,r]
is
\begin{equation}
M (r;r_0,\rho_c,z_0)=
\frac{1}{3}\,\rho_{{c}}{r_{{0}}}^{3}+{\frac {\rho_{{c}}{z_{{0}}}^{2} \left(r -r
_{{0}} \right) }{ \left( \cos \left( \theta \right)  \right) ^{2}}}
\quad.
\nonumber
\label{massinversesquare}
\end{equation}
The positive solution of equation (\ref{eqnvelocity})
gives the velocity as a function 
of the radius:
\begin{equation}
v(r;r_0,z_0) =
{\frac {\sqrt { \left( {r_{{0}}}^{3} \left( \cos \left( \theta
 \right)  \right) ^{2}+3\,{z_{{0}}}^{2}r-3\,{z_{{0}}}^{2}r_{{0}}
 \right) r_{{0}}}\cos \left( \theta \right) {\it v0}\,r_{{0}}}{{r_{{0}
}}^{3} \left( \cos \left( \theta \right)  \right) ^{2}+3\,{z_{{0}}}^{2
}r-3\,{z_{{0}}}^{2}r_{{0}}}}
\quad.
\label{vfirst}
\end{equation}
The differential equation which models the energy conservation
is 
\begin{equation}
\frac{1}{2}\, \left( \frac{1}{3}\,\rho_{{c}}{r_{{0}}}^{3}+{\frac {\rho_{{c}}{z_{{0}}}^
{2} \left(  r \left( t \right) -r_{{0}} \right) }{ \left( \cos \left( 
\theta \right)  \right) ^{2}}} \right)  \left( {\frac {\rm d}{{\rm d}t
}}r \left( t \right)  \right) ^{2}
-\frac{1}{6}\,\rho_{{c}}{r_{{0}}}^{3}{v_{{0}
}}^{2}=0
\quad.
\end{equation}
The analytical solution of the above differential equation
is 
\begin{eqnarray}
r(t;z0,r_0,\theta)
=
\nonumber \\
\frac{
-r_{{0}} \Bigg( - \left( \cos \left( \theta \right)  \right) ^{2/3}
 \left( 9\,{z_{{0}}}^{2} \left( t-t_{{0}} \right) v_{{0}}+2\,{r_{{0}}}
^{3} \left( \cos \left( \theta \right)  \right) ^{2} \right) ^{2/3}
\sqrt [3]{2}+2\,{r_{{0}}}^{2} \left( \cos \left( \theta \right) 
 \right) ^{2}-6\,{z_{{0}}}^{2} \Bigg ) 
}{6\,{z_{{0}}}^{2}}
\quad.
\label{rtinversesquare}
\end{eqnarray}
The above inverse square  profile for density 
satisfies the basic symmetries as outlined
in equations (\ref{symmetries1}) and (\ref{symmetries2}).

\subsection{Back reaction for the inverse square profile }

\begin{figure*}
\begin{center}
\includegraphics[width=5cm]{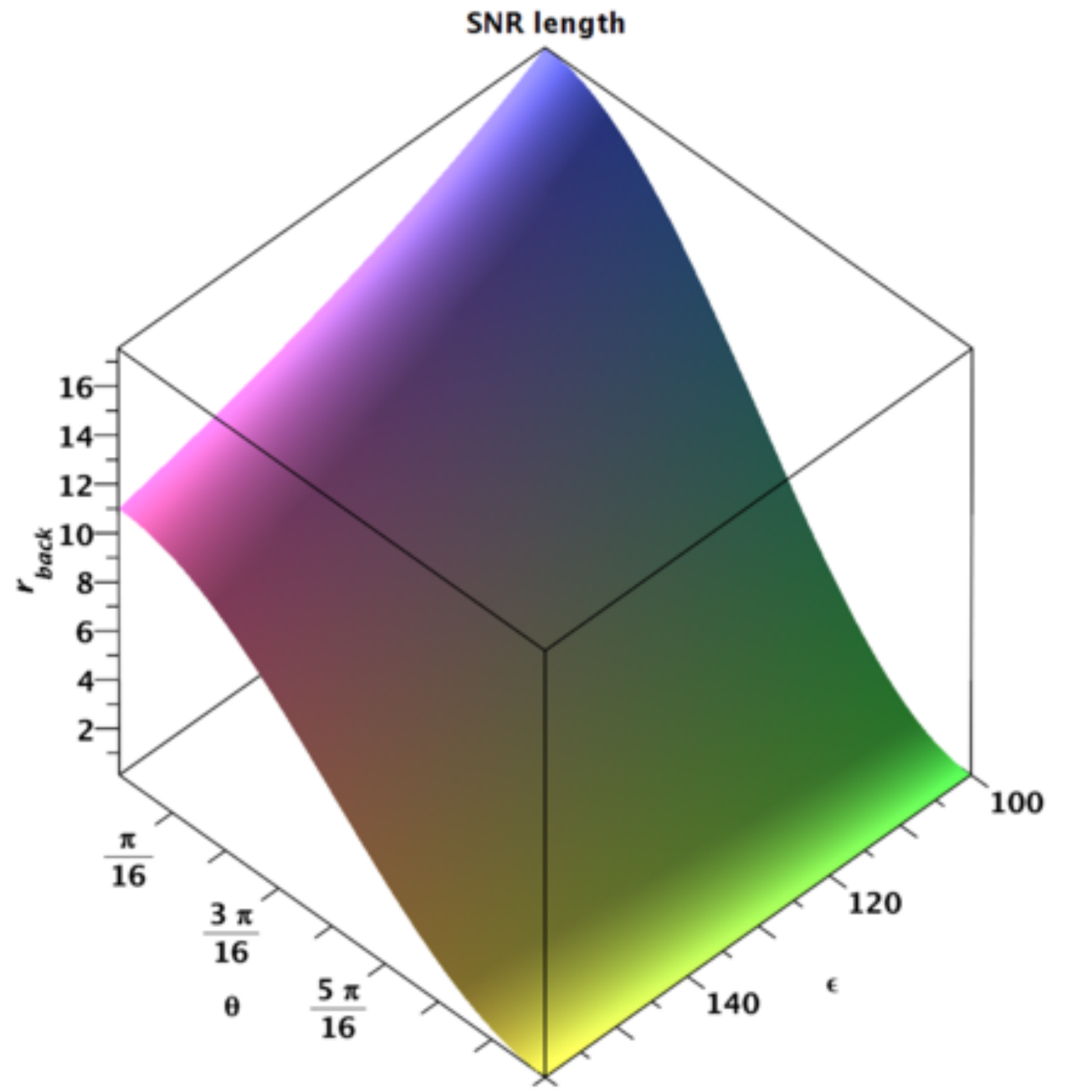}
\end {center}
\caption
{
A 2D map of $r_{back}(r_0,z_0,\theta,\epsilon)$ in pc.
The parameters are  
$r_0=0.098\,\mbox{pc}$ and $z_0=1.5\,10^{-4}\,pc$.
}
\label{snrlenght}
    \end{figure*}

The radiative losses per unit length  
are assumed to scale as 
\begin{equation}
- 4\,\epsilon\,\rho_s\,{v}^{2}{r}^{2}
\quad,
\end{equation}
where $\epsilon$ is a constant and $rho_s$ is the density
in the thin advancing layer which is $4\,\rho$.
Inserting into the above equation  the  velocity to first order
as  given by equation~(\ref{vfirst}),
the radiative losses $Q(r;r_0,v_0,z_0,\epsilon,\theta)$ are
\begin{equation}
Q(r;r_0,v_0,z_0,\epsilon,\theta)=
-
4\,{\frac {\epsilon\,{\it \rho\_c}\,{z_{{0}}}^{2}{{\it r0}}^{3}{v_{{0}}
}^{2}}{{{\it r0}}^{3} \left( \cos \left( \theta \right)  \right) ^{2}+
3\,{z_{{0}}}^{2}r-3\,{z_{{0}}}^{2}{\it r0}}}
\quad.
\label{losses}
\end{equation}
The sum of the radiative  losses between $r_0$ and $r$ 
is given by the integral
\begin{eqnarray}
L(r;r_0,v_0,z_0,\epsilon,\theta)=\int_{r_0}^r  
Q(r;r_0,v_0,z_0,\epsilon,\theta) dr
= \nonumber \\
4/3\,\epsilon\,{\it \rho\_c}\,{r_{{0}}}^{3}{v_{{0}}}^{2} \left( \ln 
 \left( {r_{{0}}}^{3} \left( \cos \left( \theta \right)  \right) ^{2}+
3\,{z_{{0}}}^{2} \left( -r_{{0}}+r \right)  \right) -\ln  \left( {r_{{0
}}}^{3} \left( \cos \left( \theta \right)  \right) ^{2} \right) 
 \right) 
\quad.
\label{classiclosses}
\end{eqnarray}
The  conservation of energy  in  the presence  
of  the back reaction due to the radiative losses
is 
\begin{eqnarray}
1/2\,{\frac {{v}^{2} \left( 1/3\,{r_{{0}}}^{3} \left( \cos \left( 
\theta \right)  \right) ^{2}+{z_{{0}}}^{2} \left( -r_{{0}}+r \right) 
 \right) {\it \rho\_c}}{ \left( \cos \left( \theta \right)  \right) ^{2
}}}+
\nonumber  \\
4/3\,\epsilon\,{\it \rho\_c}\,{r_{{0}}}^{3}{v_{{0}}}^{2}\ln 
 \left( {r_{{0}}}^{3} \left( \cos \left( \theta \right)  \right) ^{2}+
3\,{z_{{0}}}^{2} \left( -r_{{0}}+r \right)  \right) 
\nonumber \\
-4/3\,\epsilon\,{
\it \rho\_c}\,{r_{{0}}}^{3}{v_{{0}}}^{2} \left( 3\,\ln  \left( r_{{0}}
 \right) +\ln  \left(  \left( \cos \left( \theta \right)  \right) ^{2}
 \right)  \right) =1/6\,{\it \rho\_c}\,{r_{{0}}}^{3}{v_{{0}}}^{2}
\quad. 
\label{eqnenergyback}
\end{eqnarray}
The  analytical solution for the velocity to 
second order, $v_c(r;r_0,v_0,z_0,\epsilon,\theta)$, 
is
\begin{equation}
v_c(r;r_0,v_0,z_0,\epsilon,\theta)=
{\frac {\sqrt {-Ar_{{0}} \left( 8\,\ln  \left( A \right) \epsilon-8\,
\ln  \left( {r_{{0}}}^{3} \left( \cos \left( \theta \right)  \right) ^
{2} \right) \epsilon-1 \right) }v_{{0}}\cos \left( \theta \right) r_{{0
}}}{A}}
\label{vcorrected}
\quad,
\end{equation}
where 
\begin{equation}
A={r_{{0}}}^{3} \left( \cos \left( \theta \right)  \right) ^{2}+3\,{z_{{0
}}}^{2}r-3\,{z_{{0}}}^{2}r_{{0}}
\quad.
\end{equation}
The  inclusion  of the back reaction  allows the evaluation of the 
SRS's maximum length,  $r_{back}(r_0,z_0,\theta,\epsilon)$,  
which can be derived 
by setting this velocity to zero:
\begin{equation}
r_{back}(r_0,z_0,\theta,\epsilon) 
= 
\frac
{
{{\rm e}^{1/8\,{\frac {8\,\ln  \left( {r_{{0}}}^{3} \left( \cos
 \left( \theta \right)  \right) ^{2} \right) \epsilon+1}{\epsilon}}}}-
{r_{{0}}}^{3} \left( \cos \left( \theta \right)  \right) ^{2}+3\,{z_{{0
}}}^{2}r_{{0}}
}
{
3\,{z_{{0}}}^{2}
}
\quad.
\label{finitelength}
\end{equation}
Figure \ref{snrlenght} shows
the SRS's maximum length   as a function of the constant of 
conversion $\epsilon$  and  the polar angle $\theta$.
% figure   snrlenght
% figure   snrlenght
The above figure shows that the SNR cannot reach an
infinite length at an infinite time but 
has a finite lifetime and length.

\subsection{A power law profile of the density}
\label{section_powerlaw}

The medium 
is assumed to scale as 
\begin{equation}
 \rho (r;r_0)  = \{ \begin{array}{ll}
            \rho_c                      & \mbox {if $r \leq r_0 $ } \\
            \rho_c (\frac{r_0}{z})^{\alpha}    & \mbox {if $r >     r_0 $}
            \end{array}
\label{piecewisepower}
\quad, 
\end{equation}
where $\alpha$ is a positive real number.
The total mass $M(r;r_0,\rho_c\alpha)$
swept in the interval [0,r]
is
\begin{equation}
M (r;r_0,\rho_c,z_0,\alpha)=
\frac{1}{3}\,\rho_{{c}}{r_{{0}}}^{3}-{\frac {\rho_{{c}} \left( {r}^{-\alpha+3}
{z_{{0}}}^{\alpha} \left(  \left( \cos \left( \theta \right)  \right) 
^{-1} \right) ^{\alpha}-{r_{{0}}}^{-\alpha+3}{z_{{0}}}^{\alpha}
 \left(  \left( \cos \left( \theta \right)  \right) ^{-1} \right) ^{
\alpha} \right) }{\alpha-3}}
\quad.
\nonumber
\label{massalpha}
\end{equation}
The velocity $v(r;r_0,v_0,\alpha)$ as a function of the radius is
\begin{eqnarray}
v(r;r_0,v_0,\alpha)= 
\nonumber \\
-{\frac {\sqrt {-3\,r_{{0}} \left( \alpha-3 \right)  \left( -{r_{{0}}}
^{-\alpha+3}{z_{{0}}}^{\alpha} \left(  \left( \cos \left( \theta
 \right)  \right) ^{-1} \right) ^{\alpha}+{r}^{-\alpha+3}{z_{{0}}}^{
\alpha} \left(  \left( \cos \left( \theta \right)  \right) ^{-1}
 \right) ^{\alpha}-\frac{1}{3}\,{r_{{0}}}^{3} \left( \alpha-3 \right) 
 \right) }{\it v0}\,r_{{0}}}{3\,{r_{{0}}}^{-\alpha+3}{z_{{0}}}^{\alpha
} \left(  \left( \cos \left( \theta \right)  \right) ^{-1} \right) ^{
\alpha}-3\,{r}^{-\alpha+3}{z_{{0}}}^{\alpha} \left(  \left( \cos
 \left( \theta \right)  \right) ^{-1} \right) ^{\alpha}+{r_{{0}}}^{3}
 \left( \alpha-3 \right) }}
\quad.
\end{eqnarray}
The differential equation which governs the motion is
\begin{eqnarray}
\frac{1}{2}\, \left( \frac{1}{3}\,\rho_{{c}}{r_{{0}}}^{3}-{\frac {\rho_{{c}} \left( 
 \left( r \left( t \right)  \right) ^{-\alpha+3}{z_{{0}}}^{\alpha}
 \left(  \left( \cos \left( \theta \right)  \right) ^{-1} \right) ^{
\alpha}-{r_{{0}}}^{-\alpha+3}{z_{{0}}}^{\alpha} \left(  \left( \cos
 \left( \theta \right)  \right) ^{-1} \right) ^{\alpha} \right) }{
\alpha-3}} \right)  \left( {\frac {\rm d}{{\rm d}t}}r \left( t
 \right)  \right) ^{2}
\nonumber \\
-\frac{1}{6}\,\rho_{{c}}{r_{{0}}}^{3}{v_{{0}}}^{2}=0
\quad.
\label{diffequationalpha}
\end{eqnarray}
The above differential equation  does not have an analytical solution;
a Taylor expansion of order 3 covers a limited range in time
\begin{equation}
r(t;r_0,v_0,z_0,t_0,\alpha) =
r_{{0}}+v_{{0}} \left( t-t_{{0}} \right) -\frac{3}{4}\,{r_{{0}}}^{-1-\alpha}{v
_{{0}}}^{2}{z_{{0}}}^{\alpha} \left(  \left( \cos \left( \theta
 \right)  \right) ^{-1} \right) ^{\alpha} \left( t-t_{{0}} \right) ^{2
}
\quad.
\end{equation}
The above power law  profile for the density 
satisfies the basic symmetries as outlined
in equations (\ref{symmetries1}) and (\ref{symmetries2}).

\subsection{An exponential profile of the density }
\label{section_exponential}

The medium 
is assumed to scale as 
\begin{equation}
 \rho (r;r_0,z_0)  = \{ \begin{array}{ll}
            \rho_c                         & \mbox {if $r \leq r_0 $ } \\
            \rho_c \exp(-\frac{z}{z_0})    & \mbox {if $r >     r_0 $}
            \end{array}
\label{piecewiseexp}
\quad. 
\end{equation}

The total mass $ M(r;r_0,\rho_c,z_0) $
swept in the interval [0,r]
is
\begin{eqnarray}
M (r;r_0,\rho_c,z_0)=
\frac
{1}
{ 
3\, \left( \cos \left( \theta \right)  \right) ^{3}
}
\rho_{{c}} \Bigg( {r_{{0}}}^{3} \left( \cos \left( \theta \right) 
 \right) ^{3}+3\, \left( \cos \left( \theta \right)  \right) ^{2}{
{\rm e}^{-{\frac {r_{{0}}\cos \left( \theta \right) }{z_{{0}}}}}}{r_{{0
}}}^{2}z_{{0}}
\nonumber \\
-3\, \left( \cos \left( \theta \right)  \right) ^{2}{
{\rm e}^{-{\frac {r\cos \left( \theta \right) }{z_{{0}}}}}}{r}^{2}z_{{0
}}+6\,\cos \left( \theta \right) {{\rm e}^{-{\frac {r_{{0}}\cos
 \left( \theta \right) }{z_{{0}}}}}}r_{{0}}{z_{{0}}}^{2}-6\,\cos
 \left( \theta \right) {{\rm e}^{-{\frac {r\cos \left( \theta \right) 
}{z_{{0}}}}}}r{z_{{0}}}^{2}+6\,{{\rm e}^{-{\frac {r_{{0}}\cos \left( 
\theta \right) }{z_{{0}}}}}}{z_{{0}}}^{3}
\nonumber \\
-6\,{{\rm e}^{-{\frac {r\cos
 \left( \theta \right) }{z_{{0}}}}}}{z_{{0}}}^{3} \Bigg ) 
\quad.
\end{eqnarray}
The velocity $v(r;r_0,v_0)$ as a function of the radius is
\begin{eqnarray}
v(r;r_0,v_0)= 
\nonumber \\
\frac 
{
1
}
{
3\, \left( \cos \left( \theta \right)  \right) ^{3}
}
\Bigg( {r_{{0}}}^{3} \left( \cos \left( \theta \right)  \right) ^{3}+
3\, \left( \cos \left( \theta \right)  \right) ^{2}{{\rm e}^{-{\frac {
r_{{0}}\cos \left( \theta \right) }{z_{{0}}}}}}{r_{{0}}}^{2}z_{{0}}-3
\, \left( \cos \left( \theta \right)  \right) ^{2}{{\rm e}^{-{\frac {r
\cos \left( \theta \right) }{z_{{0}}}}}}{r}^{2}z_{{0}}
\nonumber \\
+6\,\cos \left( 
\theta \right) {{\rm e}^{-{\frac {r_{{0}}\cos \left( \theta \right) }{
z_{{0}}}}}}r_{{0}}{z_{{0}}}^{2}-6\,\cos \left( \theta \right) {{\rm e}
^{-{\frac {r\cos \left( \theta \right) }{z_{{0}}}}}}r{z_{{0}}}^{2}+6\,
{{\rm e}^{-{\frac {r_{{0}}\cos \left( \theta \right) }{z_{{0}}}}}}{z_{
{0}}}^{3}-6\,{{\rm e}^{-{\frac {r\cos \left( \theta \right) }{z_{{0}}}
}}}{z_{{0}}}^{3} \Bigg) \rho_{{c}}
\quad.
\end{eqnarray}
The differential equation which governs the motion is
\begin{eqnarray}
\frac
{
1
}
{
6\, \left( \cos \left( \theta \right)  \right) ^{3}
}
3\, \Bigg( -2\, \left( \frac{1}{2}\, \left( \cos \left( \theta \right) 
 \right) ^{2} \left( r \left( t \right)  \right) ^{2}+r \left( t
 \right) \cos \left( \theta \right) z_{{0}}+{z_{{0}}}^{2} \right) 
 \left( {\frac {\rm d}{{\rm d}t}}r \left( t \right)  \right) ^{2}z_{{0
}}{{\rm e}^{-{\frac {r \left( t \right) \cos \left( \theta \right) }{z
_{{0}}}}}}
\nonumber \\
+z_{{0}} \left( {\frac {\rm d}{{\rm d}t}}r \left( t \right) 
 \right) ^{2} \left(  \left( \cos \left( \theta \right)  \right) ^{2}{
r_{{0}}}^{2}+2\,r_{{0}}z_{{0}}\cos \left( \theta \right) +2\,{z_{{0}}}
^{2} \right) {{\rm e}^{-{\frac {r_{{0}}\cos \left( \theta \right) }{z_
{{0}}}}}}
\nonumber \\
-\frac{1}{3}\,{r_{{0}}}^{3} \left( \cos \left( \theta \right) 
 \right) ^{3} \left( v_{{0}}-{\frac {\rm d}{{\rm d}t}}r \left( t
 \right)  \right)  \left( v_{{0}}+{\frac {\rm d}{{\rm d}t}}r \left( t
 \right)  \right)  \Bigg) \rho_{{c}}
=0
\quad.
\label{diffequationexp}
\end{eqnarray}
In the absence  of an analytical solution, we present
the Taylor expansion of order 3 for the trajectory 
\begin{equation}
r(t;r_0,v_0,z_0,t_0) =
r_{{0}}+{\it v0}\, \left( t-{\it t0} \right) -\frac{3}{4}\,{\frac {{{\it v0}}^
{2} \left( t-{\it t0} \right) ^{2}}{r_{{0}}}{{\rm e}^{-{\frac {r_{{0}}
\cos \left( \theta \right) }{z_{{0}}}}}}}
\quad.
\end{equation}
The above exponential profile for the density
satisfies the basic symmetries as outlined
in equations (\ref{symmetries1}) and (\ref{symmetries2}).

\subsection{A toroidal profile of the density }
\label{section_toroidal}
\begin{figure*}
\begin{center}
\includegraphics[width=5cm]{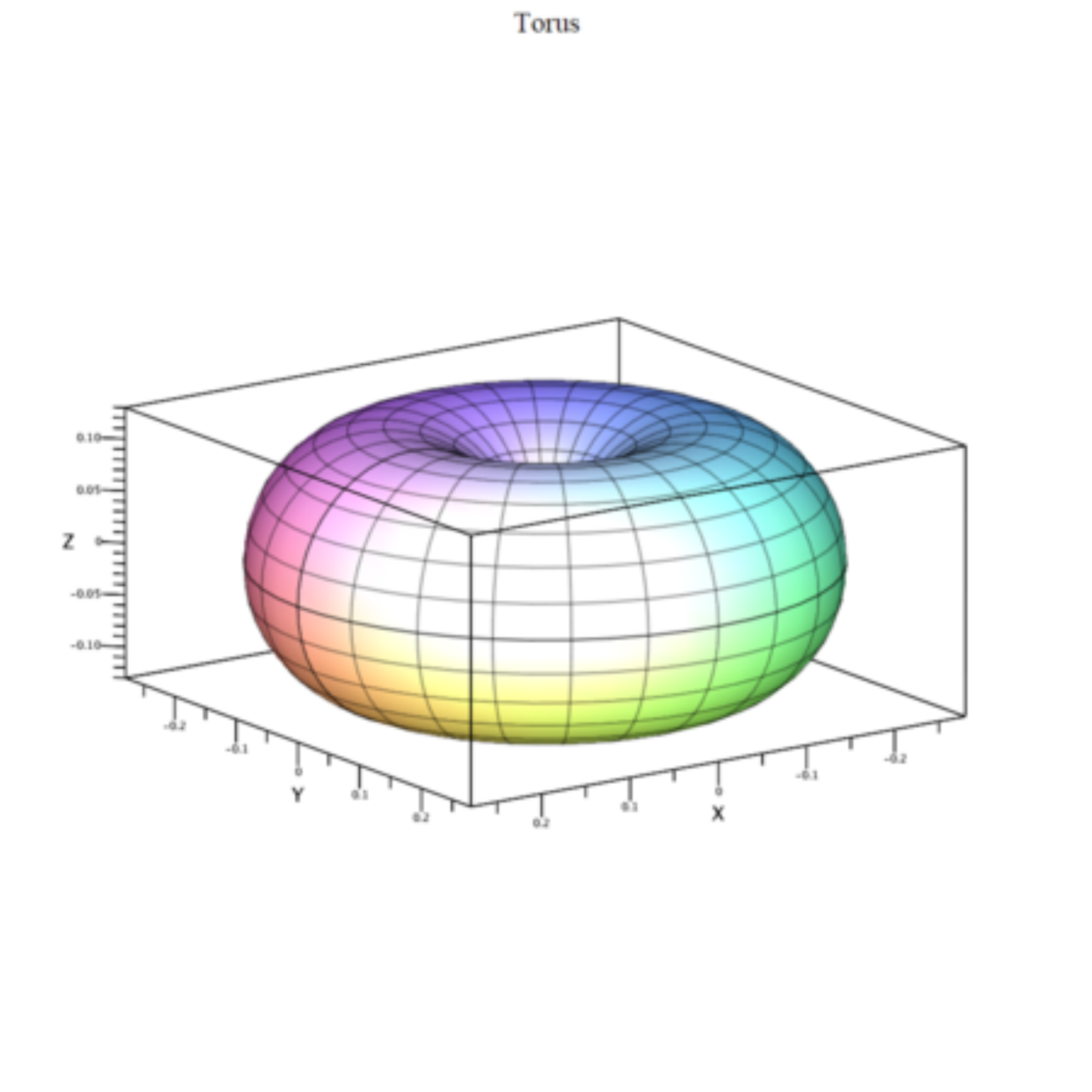}
\end {center}
\caption
{
A 3D view of a torus
}
\label{toruastro}
    \end{figure*}

\begin{figure*}
\begin{center}
\includegraphics[width=5cm]{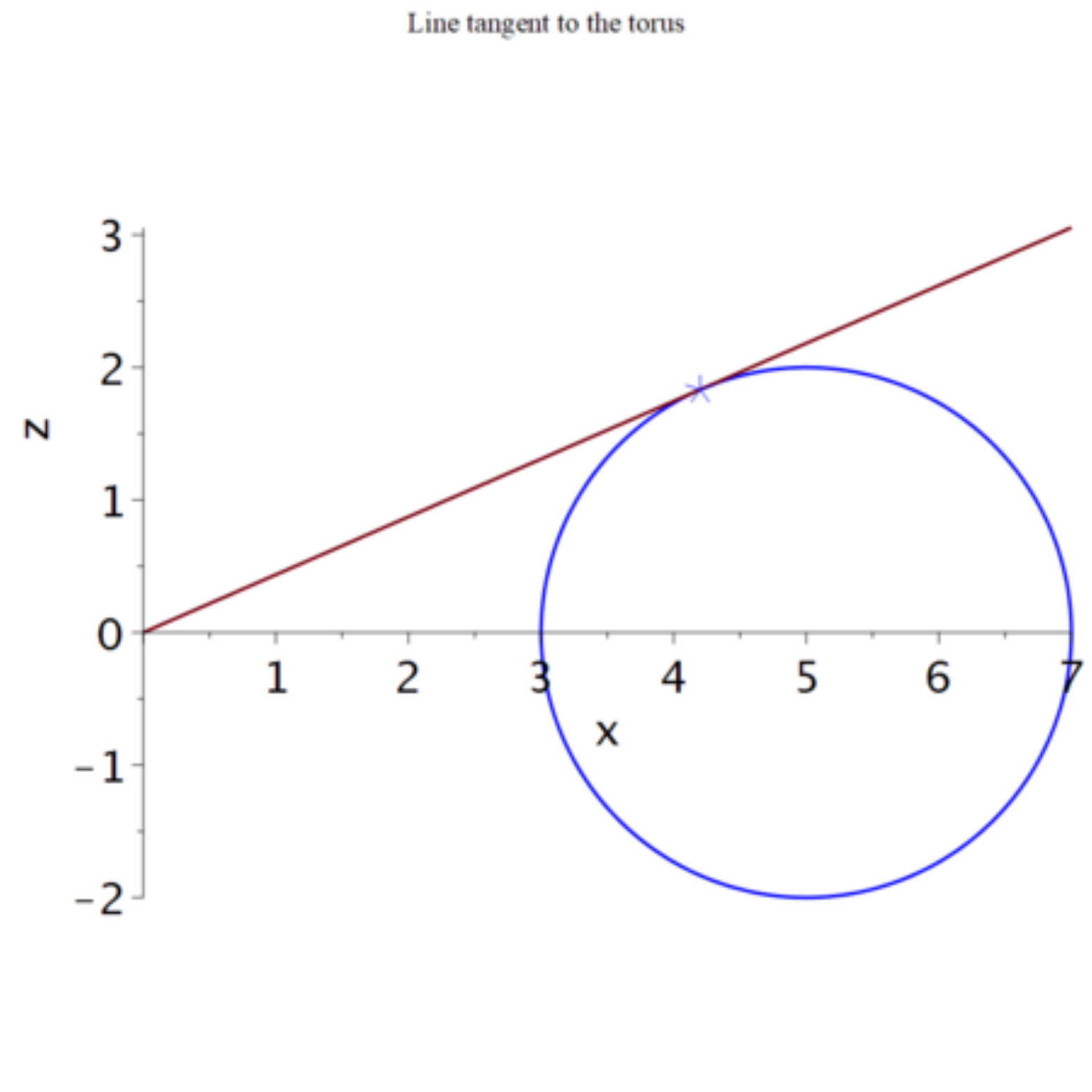}
\end {center}
\caption
{
The     straight line which intersects  the centre (red) 
and is
tangent to   the torus (blue)
in the first quadrant
}
\label{torustangent}
    \end{figure*}

\begin{figure*}
\begin{center}
\includegraphics[width=5cm]{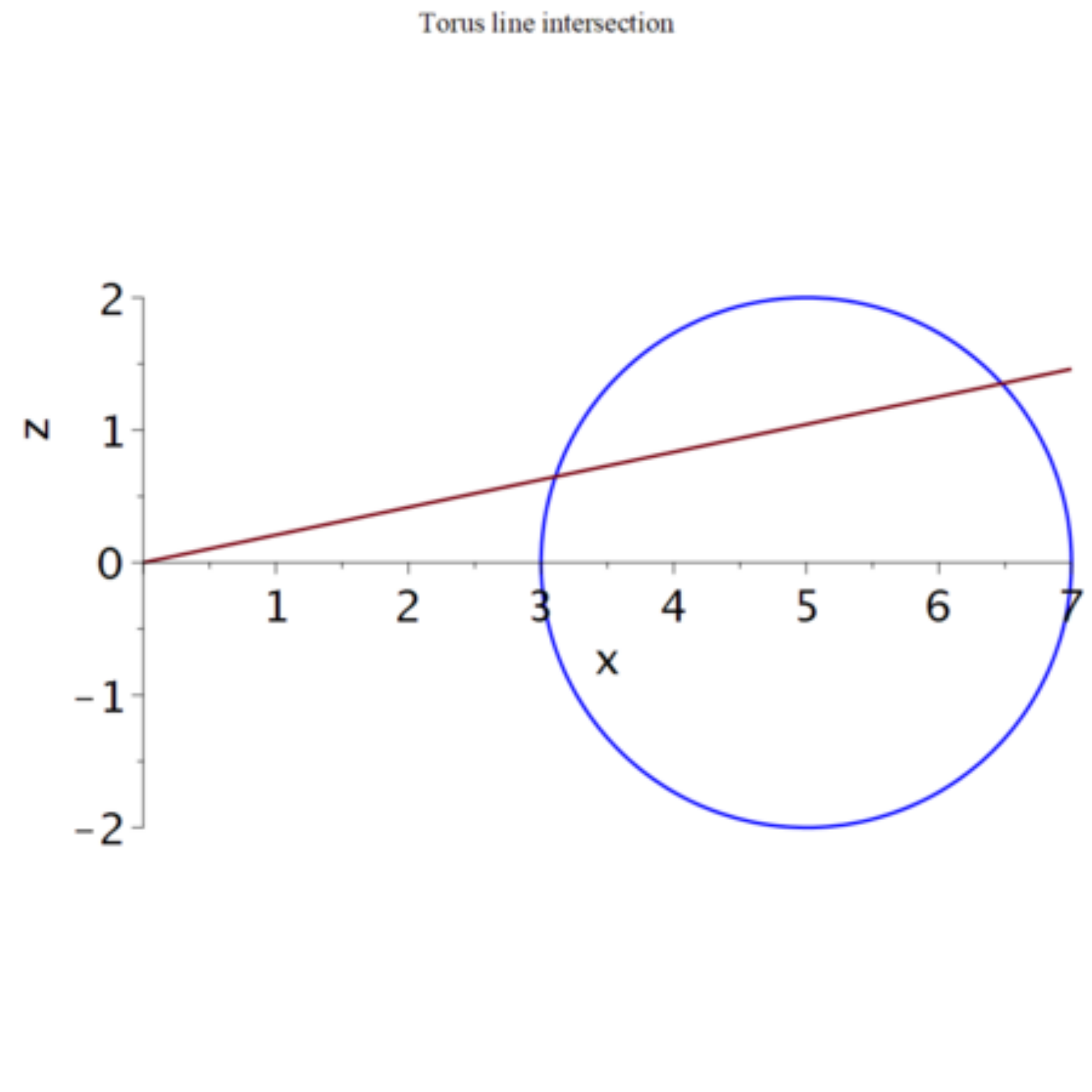}
\end {center}
\caption
{
The  intersections between a    straight line 
which intersects the centre
(red),  the torus (blue)
to the tangent point (asterisk)  
}
\label{toruslineinter}
    \end{figure*}

A torus is swept out by revolving a small circle of radius
$r_T$ about an axis lying in the same plane as the circle 
but outside it, say
at a distance of  $R_T$, 
see Figure \ref{toruastro}.
% figure   toruastro
% figure   toruastro
The above toroidal profile of the density, 
once the axis of revolution  is identified  with the polar axis, 
satisfies the basic symmetries as outlined
in equations (\ref{symmetries1}) and (\ref{symmetries2}).
In Cartesian coordinates, the torus  satisfies the equation
\begin{equation}
 \left( R_{{T}}-\sqrt {{x}^{2}+{y}^{2}} \right) ^{2}+{z}^{2}-{r_{{T}}}
^{2}=0
\quad.
\end{equation}
We now consider its intersection with the plane $y=0$
and as a consequence
the equation of the torus
is now  
\begin{equation}
 \left( R_{{T}}-x \right) ^{2}+{z}^{2}-{r_{{T}}}
^{2}=0
\quad.
\end{equation}
We now  evaluate the intersection
between the torus and 
a  straight line of equation
which crosses the centre, ($z=0,x=0$),
\begin{equation}
z = x \cot (\theta) 
\quad,
\end{equation}
where $\theta$ is the polar angle in the spherical coordinates
for the circle which represents the torus in the first quadrant.
The line touches the torus at a single point 
at the critical value of the polar angle, $\theta_{crit}$:
\begin{equation}
\theta_{crit}
=
\arctan \left( {\frac {\sqrt {{R_{{T}}}^{2}-{r_{{T}}}^{2}}}{r_{{T}}}}
 \right) 
\quad.
\end{equation}
The above angle allows us to define the following three zones.
\begin{enumerate} 
\item 
$\theta < \theta_{crit} $: the line does not intersect the torus.
\item
$\theta = \theta_{crit} $: the line is tangent to the torus at 
one degenerate single point,
with Cartesian coordinates ($x_{crit},z_{crit}$),
see Figure \ref{torustangent}.
\item
$\theta > \theta_{crit}$: the line intersects the torus 
at two points, with Cartesian coordinates ($x_1,z_1$) and 
($x_2,z_2$), see  Figure \ref{toruslineinter}.
\end{enumerate}
% figure   torustangent
% figure   torustangent
% figure   toruslineinter
% figure   toruslineinter

The Cartesian coordinates at $\theta = \theta_{crit}$
are  
\begin{subequations}
\begin{align}
x_{crit} = {\frac {{R_{{T}}}^{2}-{r_{{T}}}^{2}}{R_{{T}}}}
\\
z_{crit} = {\frac {\sqrt {{R_{{T}}}^{2}-{r_{{T}}}^{2}}r_{{T}}}{R_{{T}}}}
\quad.
\end{align}
\end{subequations}
The Cartesian coordinates of the two intersections
when  $\theta > \theta_{crit}$ are
\begin{subequations}
\begin{align}
x_1 =
\left( R_{{T}}\sin \left( \theta \right) -\sqrt {-{R_{{T}}}^{2}
 \left( \cos \left( \theta \right)  \right) ^{2}+{r_{{T}}}^{2}}
 \right) \sin \left( \theta \right) 
\\
z_1   =
\left( R_{{T}}\sin \left( \theta \right) -\sqrt {-{R_{{T}}}^{2}
 \left( \cos \left( \theta \right)  \right) ^{2}+{r_{{T}}}^{2}}
 \right) \cos \left( \theta \right) 
\\
x_2 = 
\left( R_{{T}}\sin \left( \theta \right) +\sqrt {-{R_{{T}}}^{2}
 \left( \cos \left( \theta \right)  \right) ^{2}+{r_{{T}}}^{2}}
 \right) \sin \left( \theta \right) 
\\
z_2   =
\left( R_{{T}}\sin \left( \theta \right) +\sqrt {-{R_{{T}}}^{2}
 \left( \cos \left( \theta \right)  \right) ^{2}+{r_{{T}}}^{2}}
 \right) \cos \left( \theta \right) 
\quad.
\end{align}
\end{subequations}
We now assume that the density of matter
is  $\rho_c$ outside the torus  and  $\rho_1$ 
inside the torus.
The mass swept in the third zone requires 
a careful analysis.
When  $\theta > \theta_{crit}$,
the mass swept along a line  
before the intersection with the torus, $M_I(r;\rho_c)$,
is 
\begin{equation}
M_I(r;\rho_c) =\frac{1}{3}\,\rho_c\,{r}^{3}
\quad, 
\end{equation}
where $r$ is the momentary radius of expansion in spherical
coordinates.

The mass swept  when $r$ is inside the torus 
in the third zone, $M_{II}(r;\rho_c,\rho_1,R_T,r_T)$,
is
\begin{eqnarray}
M_{II}(r;\rho_c,\rho_1,R_T,r_T) =
\frac{1}{3}\,\rho_{{c}} \left(  \left( R_{{T}}\sin \left( \theta \right) -
\sqrt {-{R_{{T}}}^{2} \left( \cos \left( \theta \right)  \right) ^{2}+
{r_{{T}}}^{2}} \right) ^{2} \right) ^{3/2}
\nonumber \\
+\frac{1}{3}\,\rho1\, \left( {r}^{3}
- \left(  \left( R_{{T}}\sin \left( \theta \right) -\sqrt {-{R_{{T}}}^
{2} \left( \cos \left( \theta \right)  \right) ^{2}+{r_{{T}}}^{2}}
 \right) ^{2} \right) ^{3/2} \right) 
\quad.
\end{eqnarray}
The mass swept  when $r$ is outside the torus 
in the third zone is
\begin{eqnarray}
M_{III}(r;\rho_c,\rho_1,R_T,r_T) =
\frac{1}{3}\,\rho_{{c}} \left(  \left( R_{{T}}\sin \left( \theta \right) -
\sqrt {-{R_{{T}}}^{2} \left( \cos \left( \theta \right)  \right) ^{2}+
{r_{{T}}}^{2}} \right) ^{2} \right) ^{3/2}
\nonumber \\
+\frac{1}{3}\,\rho1\, \Bigg( 
 \left(  \left( R_{{T}}\sin \left( \theta \right) +\sqrt {-{R_{{T}}}^{
2} \left( \cos \left( \theta \right)  \right) ^{2}+{r_{{T}}}^{2}}
 \right) ^{2} \right) ^{3/2}
\nonumber  \\
- \left(  \left( R_{{T}}\sin \left( \theta
 \right) -\sqrt {-{R_{{T}}}^{2} \left( \cos \left( \theta \right) 
 \right) ^{2}+{r_{{T}}}^{2}} \right) ^{2} \right) ^{3/2} \Bigg) 
\nonumber  \\
+\frac{1}{3}
\,\rho_{{c}} \left( {r}^{3}- \left(  \left( R_{{T}}\sin \left( \theta
 \right) +\sqrt {-{R_{{T}}}^{2} \left( \cos \left( \theta \right) 
 \right) ^{2}+{r_{{T}}}^{2}} \right) ^{2} \right) ^{3/2} \right) 
\quad.
\end{eqnarray}
The equation of motion  is solved through the Euler method 
when the following recursive equations are solved 
\begin{subequations}
\begin{align}
r_{n+1} = r_n + v_n \Delta t
\label{recursive1}   
 \\
v_{n+1} = v_n 
\Bigl (\frac {M_n(r_n)}{M_{n+1} (r_{n+1})} \Bigr )^{1/2}
\label{velocityeuler}
\quad,
\end{align}
\end{subequations}
where  $r_n$, $v_n$, $M_n$ are the temporary radius,
velocity,  and total mass, respectively,
$\Delta t $ is the time step   and $n$ is the index.
Due to the fact that the velocity is  
continuously updated, see equation (\ref{velocityeuler}), 
the method turns out to be stable.

\section{Astrophysical applications}

\label{sec_energy_applications}
\begin{figure*}
\begin{center}
\includegraphics[width=7cm]{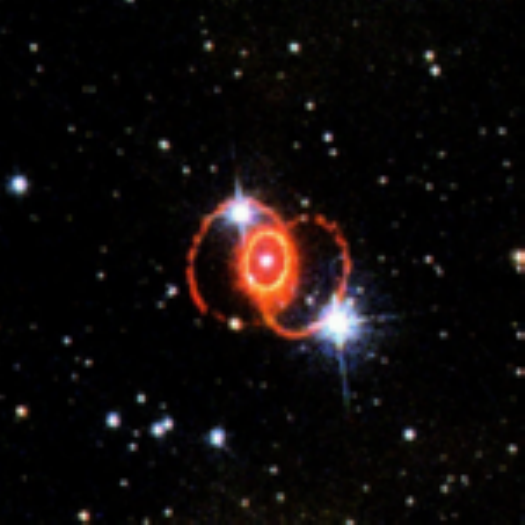}
\end {center}
\caption
{
An ST image of \sn1987a in the year 1997.
Credit 
 is  given to  
 the  Hubble Space Telescope.
}
\label{sn1987a_st}
    \end{figure*}

\begin{figure*}
\begin{center}
\includegraphics[width=7cm,angle=-90]{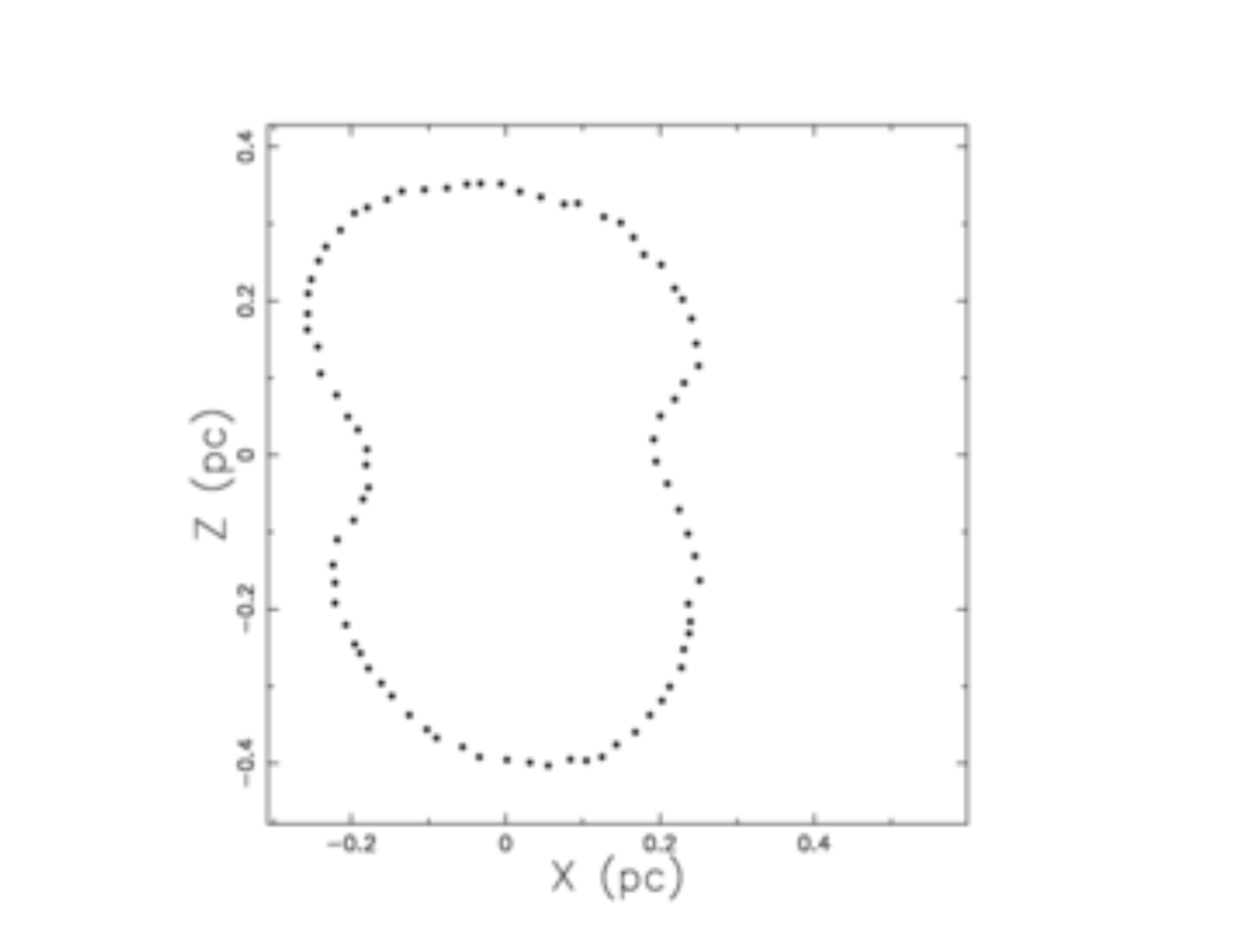}
\end {center}
\caption
{
Geometric section of \sn1987a
in the $x-z$ plane adapted by the author from Figure  5
of \cite{France2015}.
}
\label{section_obs_sn1987a}
    \end{figure*}

%inizio SN1987a
The complex structure of \sn1987a,
see ithe  Hubble Space Telescope (ST) image in Figure
\ref{sn1987a_st},
can be classified as
a torus only, a torus plus two lobes, and a torus plus 4 lobes,
see \cite{Racusin2009,McCray2017};
we therefore speak of a strong asymmetry.
% figure   sn1987a_st
% figure   sn1987a_st
The region connected with the
radius of the advancing torus is here identified
with  our equatorial
region,
in  spherical coordinates, $\theta =\frac{\pi}{2}$.
A  useful resource for calibration is the geometric  section
of \sn1987a
which is given as a sketch in Figure  5 of \cite{France2015}.
This geometric  section
was digitized  and  rotated in the $x-z$ plane
by $-40\degreezan$,  see Figure \ref{section_obs_sn1987a};
it will be the astronomical section of reference in order to test the 
simulations.

% figure   section_obs_sn1987a
\begin{figure*}
\begin{center}
\includegraphics[width=7cm]{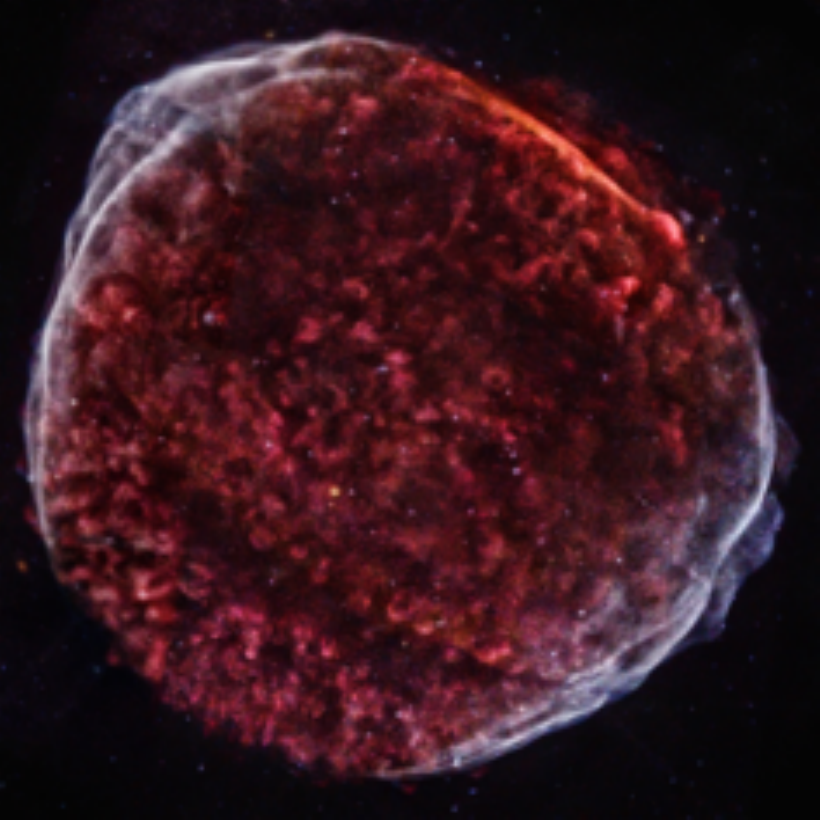}
\end {center}
\caption
{
A Chandra  X-ray (Red, Green, Blue)  image of \s1006 in the year 2013.
Credit is given to 
the Chandra X-ray Observatory.   
}
\label{SN_CHANDRA_X_2013}
    \end{figure*}

\begin{figure*}
\begin{center}
\includegraphics[width=7cm]{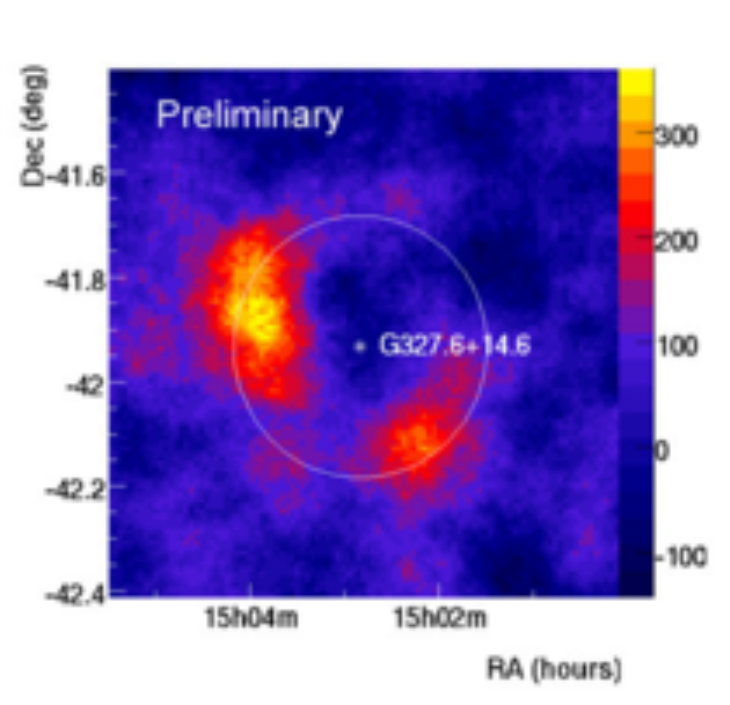}
\end {center}
\caption
{
An high energy stereoscopic system (HESS)  $\gamma$-ray image of \s1006.   
Credit is given to 
the Max-Planck-Institut f\"{u}r Kernphysik.
}
\label{SN1006_HESS}
    \end{figure*}

% figure   section_obs_sn1987a
%fine SN1987a
%inizio SN1006
\s1006   started to be
visible in 1006 AD
and currently has a
radius  of 12.2 \mbox{pc}, see \cite{Uchida2013,Katsuda2017}.
The X-shape is shown in 
Figure \ref{SN_CHANDRA_X_2013}
and  the  $\gamma$-ray  shape (100 GeV) is shown 
in Figure  \ref{SN1006_HESS}.
% figure   SN_CHANDRA_X_2013
% figure   SN_CHANDRA_X_2013

% figure   SN1006_HESS
\begin{figure*}
\begin{center}
\includegraphics[width=7cm,angle=-90]{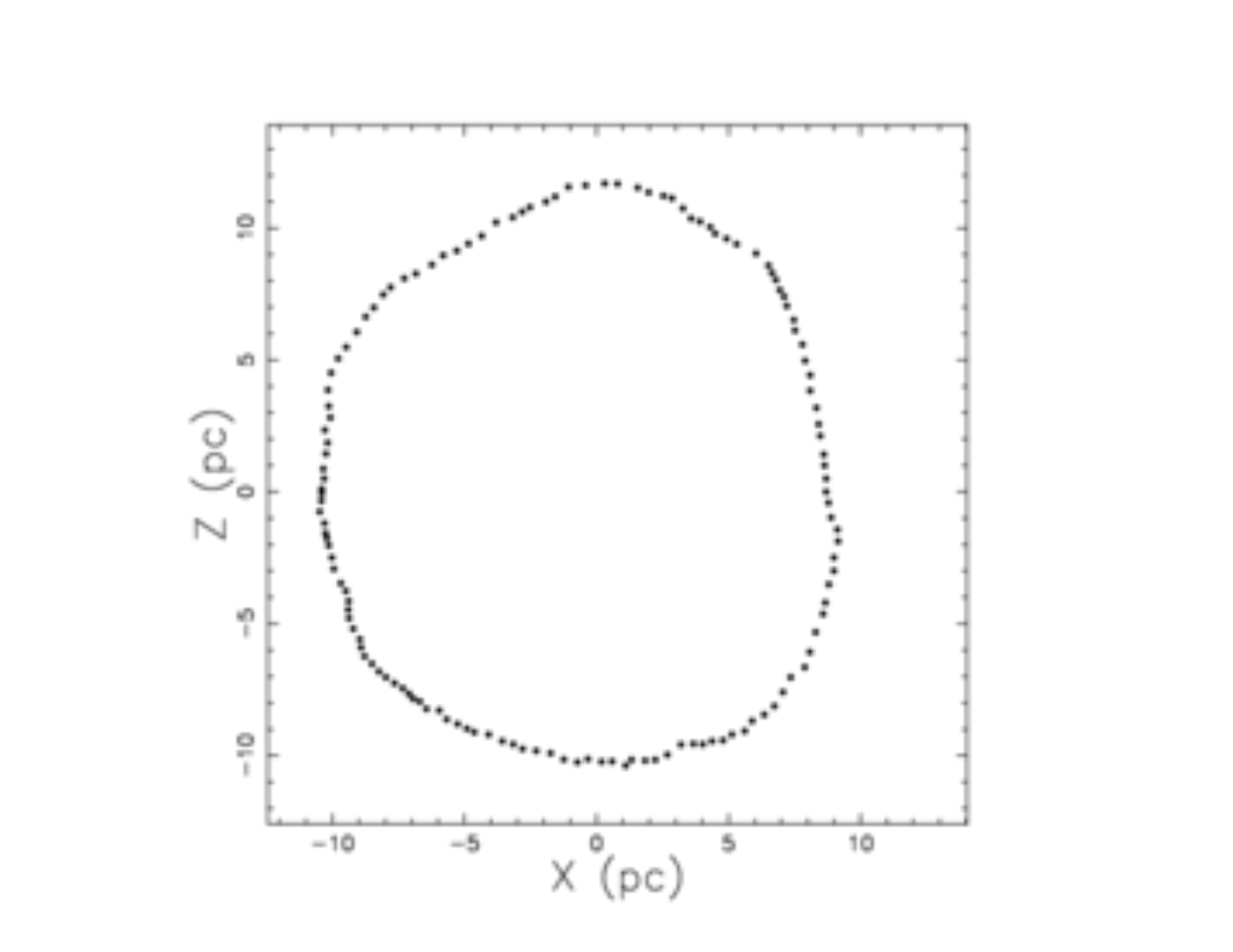}
\end {center}
\caption
{
Geometric section of \s1006
in the $x-z$ plane
adapted by the author from 
our Figure \ref{SN_CHANDRA_X_2013}. 
}
\label{sn1006_obs}
    \end{figure*}

% figure   SN1006_HESS
More precisely, on
referring to the 
above X-image, 
it   can be observed that the radius is
greatest in the north-east direction,
see also the radio map of \s1006  at 1370 MHz by
\cite {Reynolds1986},
and the X-map in the 0.4–5.0 keV band 
of Fig.~1 in \cite{Uchida2013}.
The following observed radii can be extracted:
$R_{up}=11.69$~pc in the polar  direction and $R_{eq}= 8.7$~pc in the
equatorial direction.
A geometric section of the above X-map
was digitized  and  rotated in the $x-z$ plane
by $-45\degreezan$,  see Figure \ref{sn1006_obs};
it will be the test for the simulations
% figure   sn1006_obs
% figure   sn1006_obs
Based on the previous  data we can speak of a weak asymmetry
for \s1006.
%fine di sn1006
%inizio reliability
The reliability of the models  is evaluated in terms of
the 
percentage reliability, $\epsilon_{\mathrm {obs}}$,
\begin{equation}
\epsilon_{\mathrm {obs}}  =100(1-\frac{\sum_j |r_{\mathrm {obs}}-r_{\mathrm{num}}|_j}{\sum_j
{r_{\mathrm {obs}}}_{,j}})
\quad,
\label{efficiencymany}
\end{equation}
where
$r_{\mathrm{num}}$ is the theoretical radius of the SNR,
$r_{\mathrm{obs}}$ is the observed    radius of the SNR, 
and
the  index $j$  varies  from 1 to the number of
available observations.
% fine reliability 

\subsection{Results for the inverse square profile}

\begin{figure*}
\begin{center}
\includegraphics[width=7cm,angle=-90]{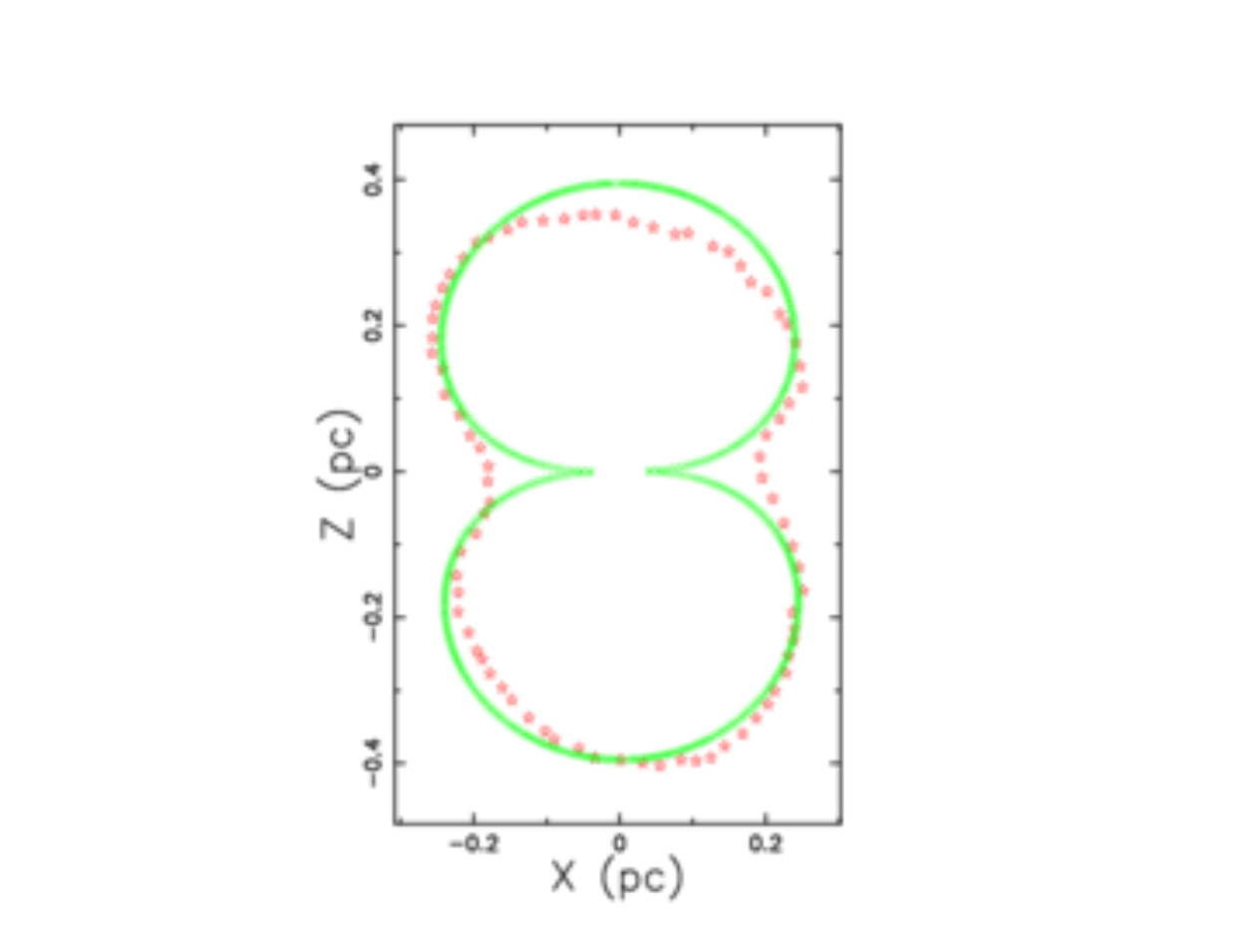}
\end {center}
\caption
{
Geometric section of \sn1987a
in the $x-z$ plane with an inverse square  profile
(green points)
and the observed profile (red stars).
The parameters
$r_0=0.001355$\ pc,
$z_0=0.0001$\ pc,
$t=27.7$\ yr,
$t_0 =0.053$ yr  and
$v_0\,=25000$ km s$^{-1}$
give
$\epsilon_{\mathrm {obs}}=91.92\%$.
}
\label{cut_invsquare_1987a}
    \end{figure*}

In the case of an inverse square  profile of
the density we have an analytical solution  as given
by equation (\ref{rtinversesquare}).
Figure \ref{cut_invsquare_1987a} displays a cut  of  \sn1987a
in the $x-z$ plane.

% figure   cut_invsquare_1987a
% figure   cut_invsquare_1987a

\begin{figure*}
\begin{center}
\includegraphics[width=7cm]{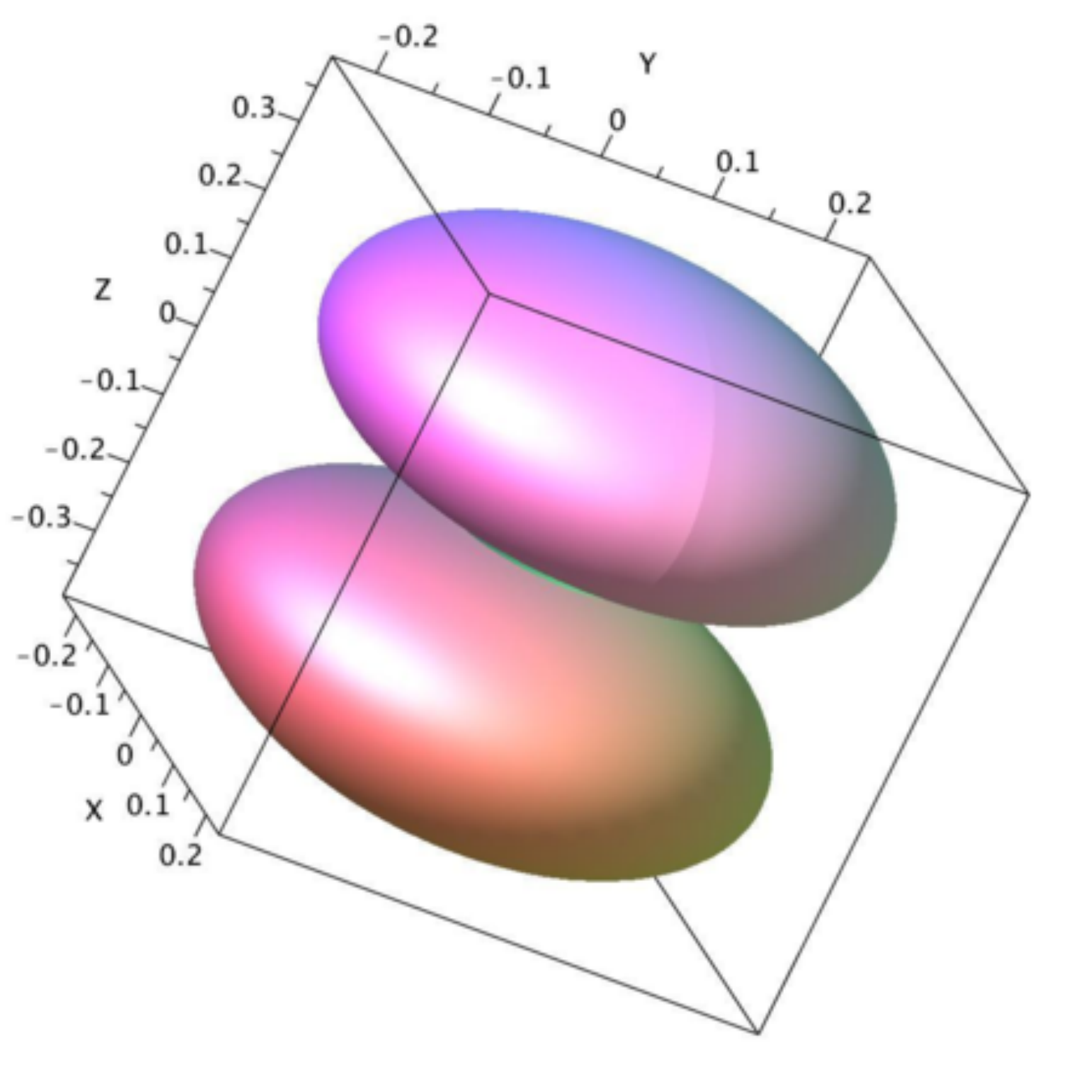}
\end {center}
\caption
{
3D surface  of  \sn1987a
with parameters as in Figure \ref{cut_invsquare_1987a} in
the framework of an 
inverse square  profile.
The three Euler angles are $\Theta=40$, $\Phi=60$ and
$ \Psi=60 $.
}
\label{3dinversesquaresn1987a}
    \end{figure*}

A rotation around  the
$z$-axis  of the previous  geometric section allows
building a 3D surface, see
Figure \ref{3dinversesquaresn1987a}.
% figure   3dinversesquaresn1987a
% figure   3dinversesquaresn1987a

Figure \ref{cut_invsquare_1987a} displays a cut  of  \sn1987a
in the $x-z$ plane.

\begin{figure*}
\begin{center}
\includegraphics[width=7cm,angle=-90]{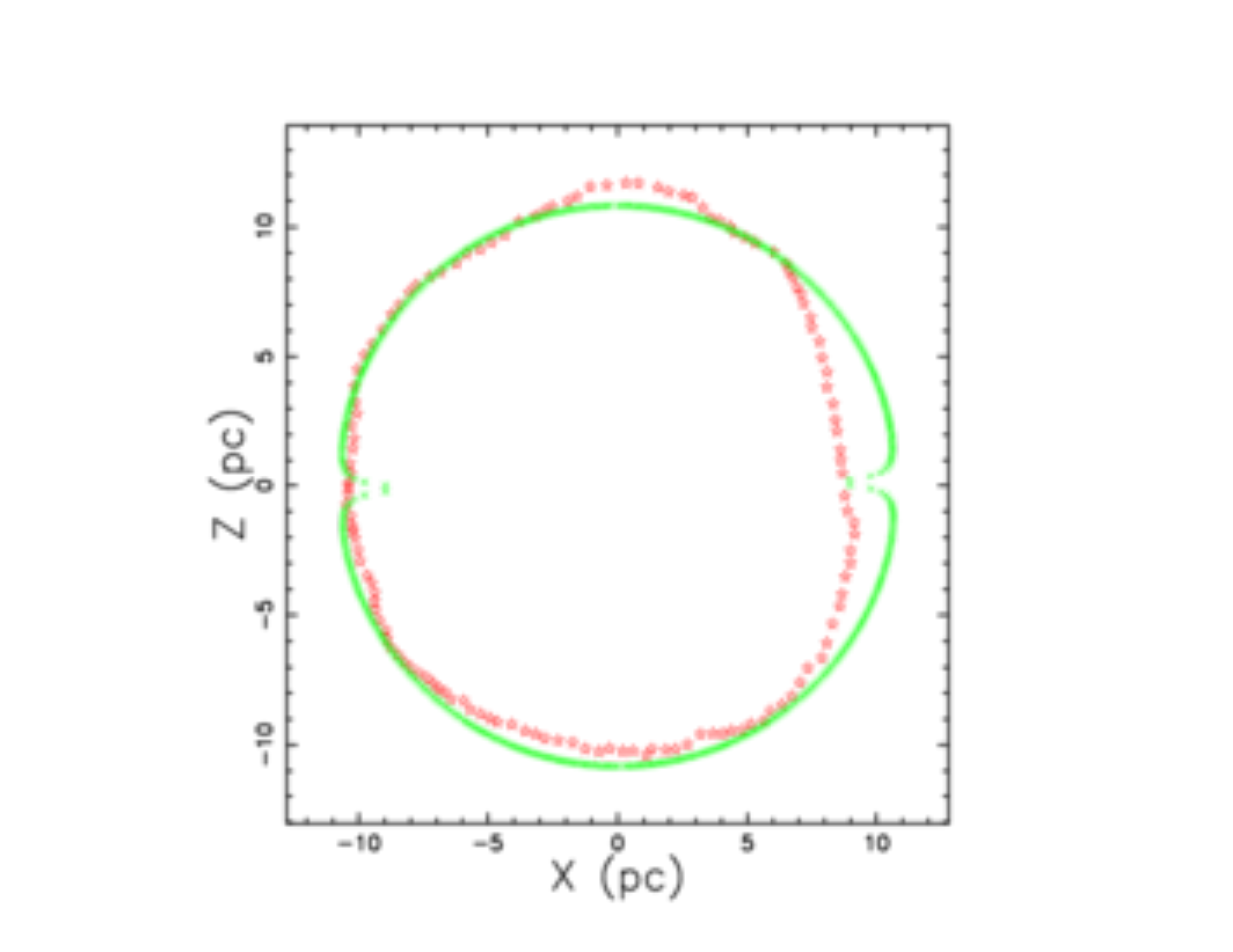}
\end {center}
\caption
{
Geometric section of \s1006
in the $x-z$ plane with an inverse square  profile
(green points)
and the observed profile (red stars).
The parameters
$r_0=0.098$  pc,
$z_0=1.5\,10^{-4}$  pc,
$t=1000$  yr,
$t_0 =9.057 $ yr  and
$v_0\,=10600$ km s$^{-1}$
give
$\epsilon_{\mathrm {obs}}=94.34\%$.
}
\label{sn1006_vera_obs_invsquare}
    \end{figure*}

\begin{figure*}
\begin{center}
\includegraphics[width=7cm]{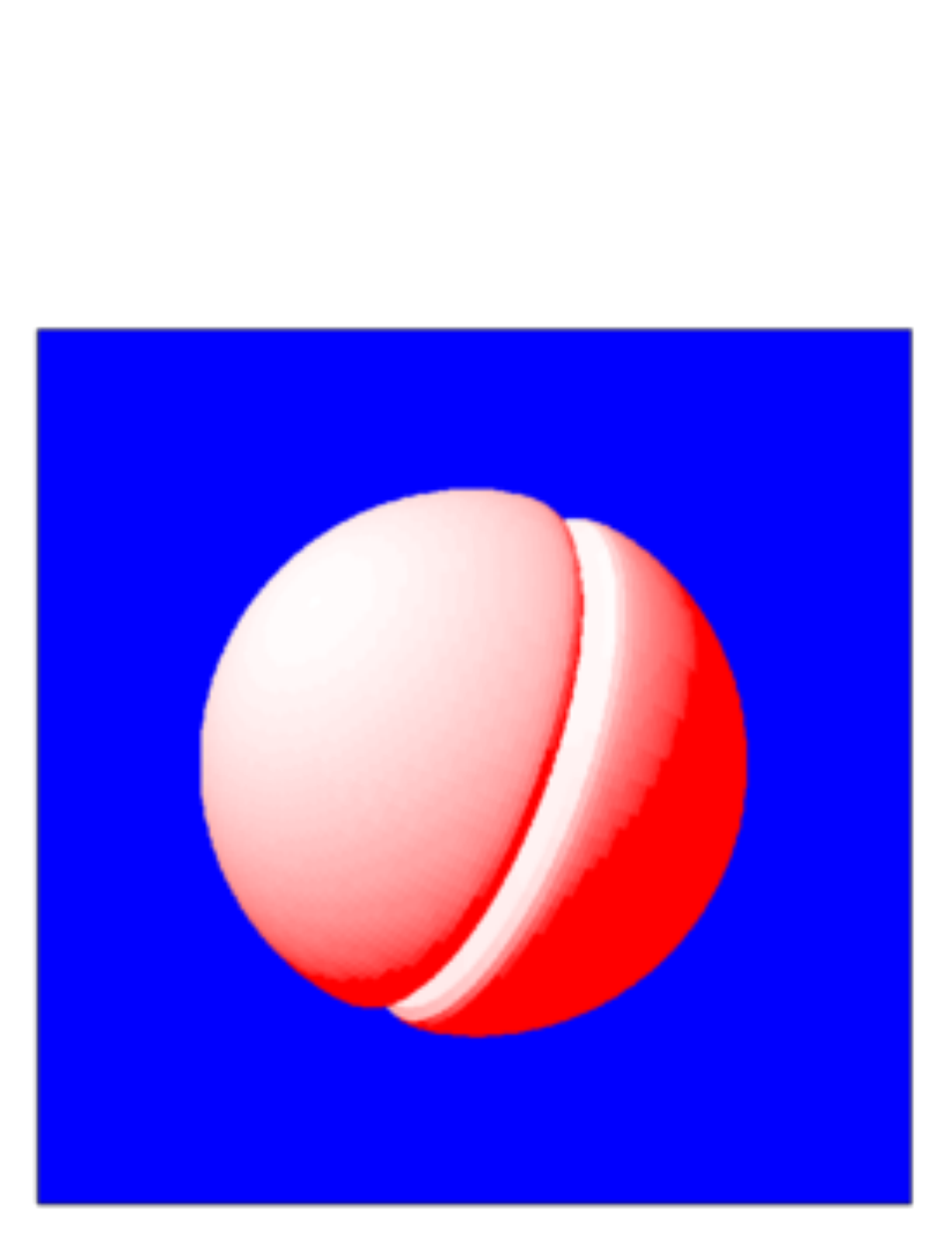}
\end {center}
\caption
{
3D surface  of  \s1006
with parameters as in Figure \ref{sn1006_vera_obs_invsquare}, 
inverse square  profile.
The three Euler angles are $\Theta=70$, $\Phi=70$ and
$ \Psi=70 $.
}
\label{3dsn1006_invsquare}
    \end{figure*}

Figure \ref{sn1006_vera_obs_invsquare} displays a cut  of  \s1006
in the $x-z$ plane
and Figure \ref{3dsn1006_invsquare} a 3D display.

% figure   sn1006_vera_obs_invsquare
% figure   sn1006_vera_obs_invsquare

% figure   3dsn1006_invsquare
% figure   3dsn1006_invsquare
The above results can be easily reproduced 
because we have an analytical expression for the trajectory.

\subsection{Results for a power profile}

\begin{figure*}
\begin{center}
\includegraphics[width=7cm,angle=-90]{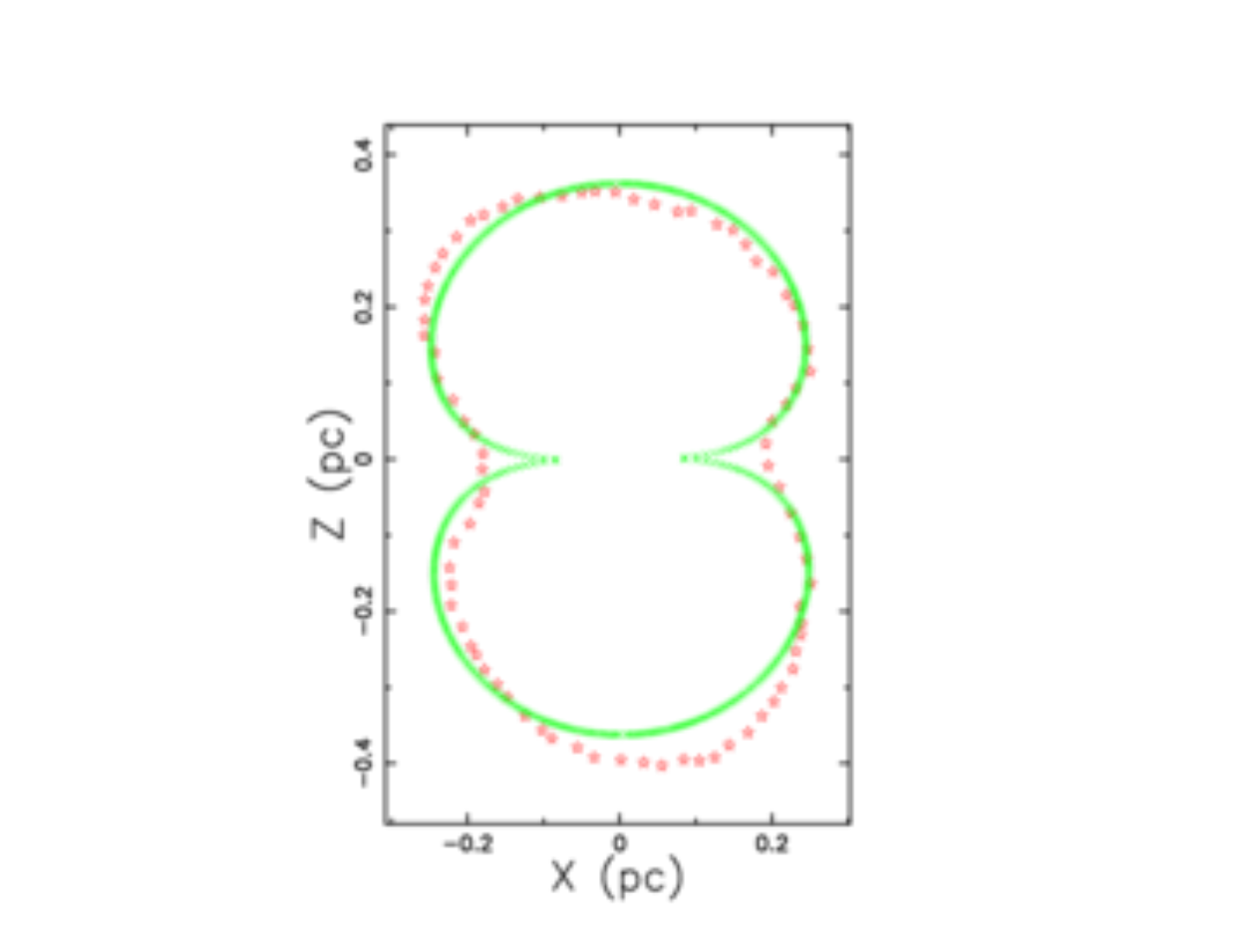}
\end {center}
\caption
{
Geometric section of \sn1987a
in the $x-z$ plane with a power law   profile
(green points)
and the observed profile (red stars).
The parameters
$\alpha=1.5$, 
$r_0=0.0038$ pc,
$z_0=0.0001$ pc,
$t=27.7$  yr,
$t_0 =0.148$ yr  and
$v_0=25000$ km s$^{-1}$
give
$\epsilon_{\mathrm {obs}}=91.92\%$.
}
\label{cut_alpha_1987a}
    \end{figure*}

\begin{figure*}
\begin{center}
\includegraphics[width=7cm,angle=-90]{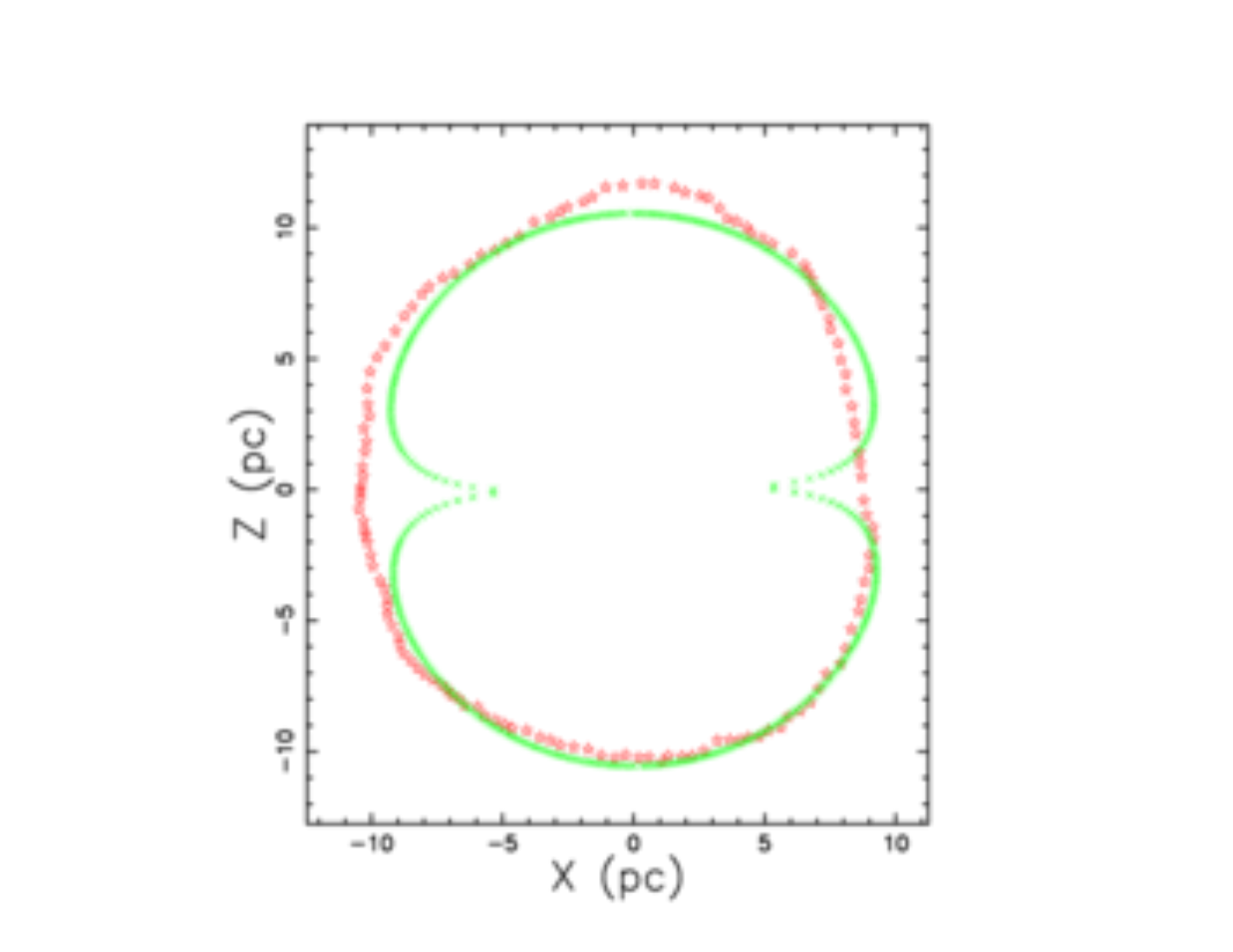}
\end {center}
\caption
{
Geometric section of \s1006
in the $x-z$ plane with a power law   profile
(green points)
and the observed profile (red stars).
The parameters
$\alpha=1.5$, 
$r_0=0.098$ pc,
$z_0=0.00015$ pc,
$t=1000$  yr,
$t_0 =9.057$ yr  and
$v_0=10600$ km s$^{-1}$
give
$\epsilon_{\mathrm {obs}}=93.562\%$.
}
\label{sn1006_vera_obs_alpha}
    \end{figure*}

In the case of a power law   profile of
the density we obtained a numerical  
solution  
for the differential equation (\ref{diffequationalpha}).
Figure \ref{cut_alpha_1987a} presentes the results 
for    \sn1987a in the $x-z$ plane
and  Figure \ref{sn1006_vera_obs_alpha} those for \s1006.

% figure   cut_alpha_1987a
% figure   cut_alpha_1987a

% figure   sn1006_vera_obs_alpha
% figure   sn1006_vera_obs_alpha

The above results 
shows that a variable
power law profile
can model the observed sections 
of the two SNRS here simulated.

\subsection{Results for an exponential profile}

\begin{figure*}
\begin{center}
\includegraphics[width=7cm,angle=-90]{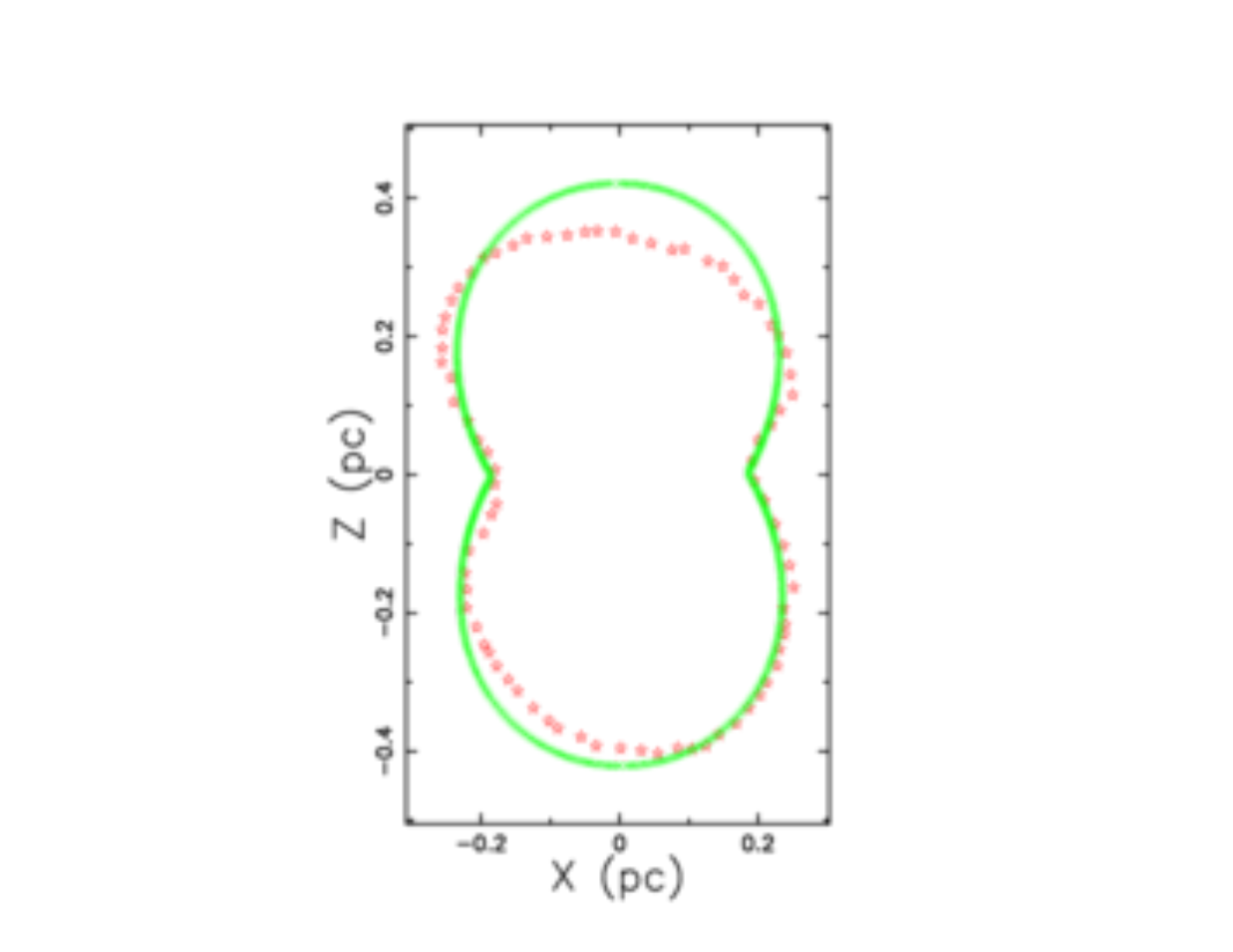}
\end {center}
\caption
{
Geometric section of \sn1987a
in the $x-z$ plane with an exponential    profile
(green points)
and the observed profile (red stars).
The parameters
$r_0=0.04 $ pc,
$z_0=0.033$ pc,
$t=27.7$  yr,
$t_0 =0.0528$ yr  and
$v_0=25000$ km s$^{-1}$
give
$\epsilon_{\mathrm {obs}}=92.45\%$.
}
\label{cut_exp_1987a}
    \end{figure*}

\begin{figure*}
\begin{center}
\includegraphics[width=7cm,angle=-90]{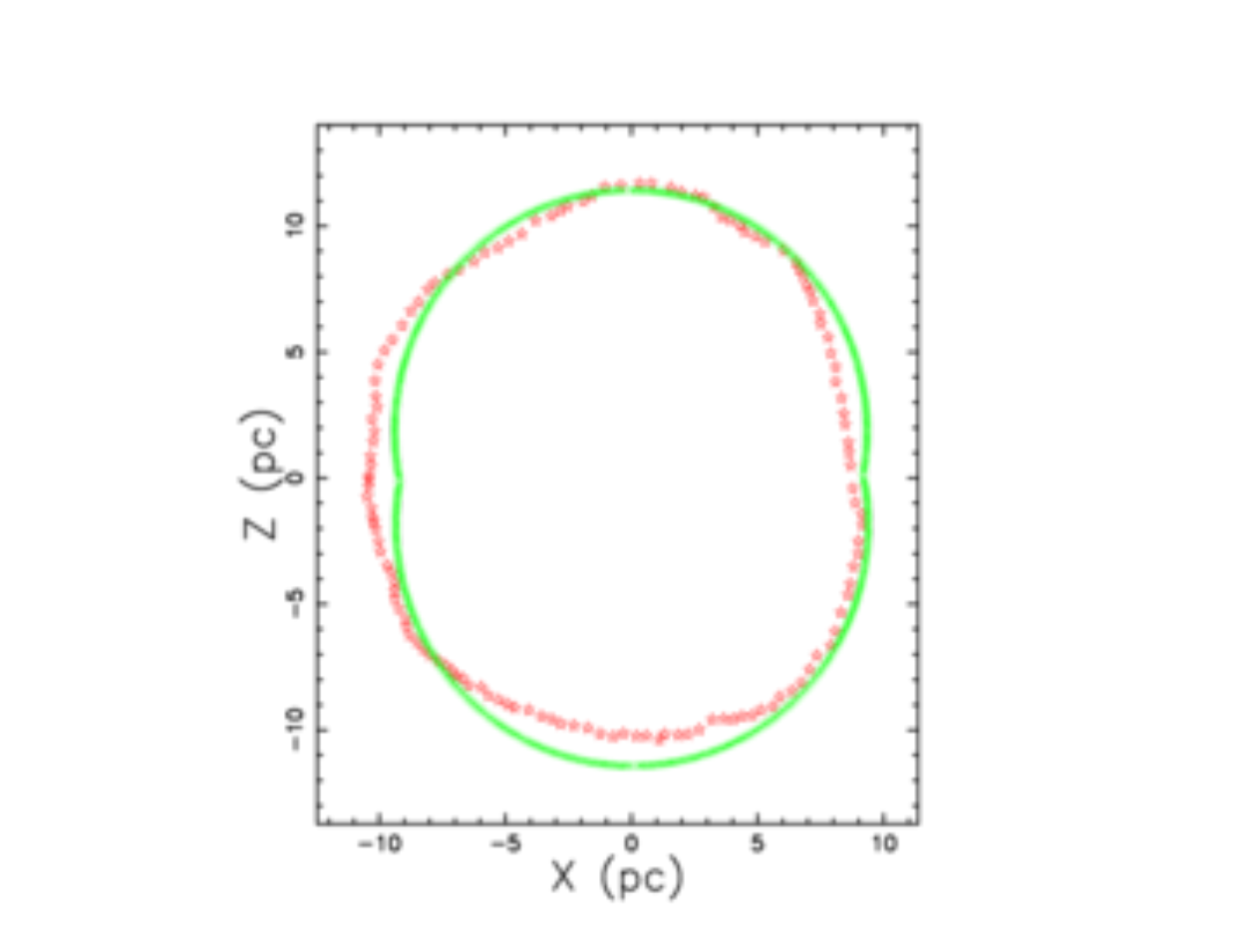}
\end {center}
\caption
{
Geometric section of \s1006
in the $x-z$ plane with an exponential  profile
(green points)
and the observed profile (red stars).
The parameters
$r_0=2.45$ pc,
$z_0=4.9 $ pc,
$t=1000$  yr,
$t_0 =0.052$ yr  and
$v_0=25000$ km s$^{-1}$
give
$\epsilon_{\mathrm {obs}}=94.26\%$.
}
\label{sn1006_vera_obs_exp}
    \end{figure*}

In the case of an exponential    profile of
the density we obtained a numerical  
solution  
for the differential equation (\ref{diffequationexp}). 
Figure \ref{cut_exp_1987a} presents the results 
for    \sn1987a in the $x-z$ plane
and  Figure \ref{sn1006_vera_obs_exp} those for \s1006.

% figure   cut_exp_1987a
% figure   cut_exp_1987a

% figure   sn1006_vera_obs_exp
% figure   sn1006_vera_obs_exp
The above results 
shows that an exponential  profile for the density
is comparable  with  the observed sections 
of the two SNRS here simulated.

\subsection{Results for a toroidal profile}

\begin{figure*}
\begin{center}
\includegraphics[width=7cm,angle=-90]{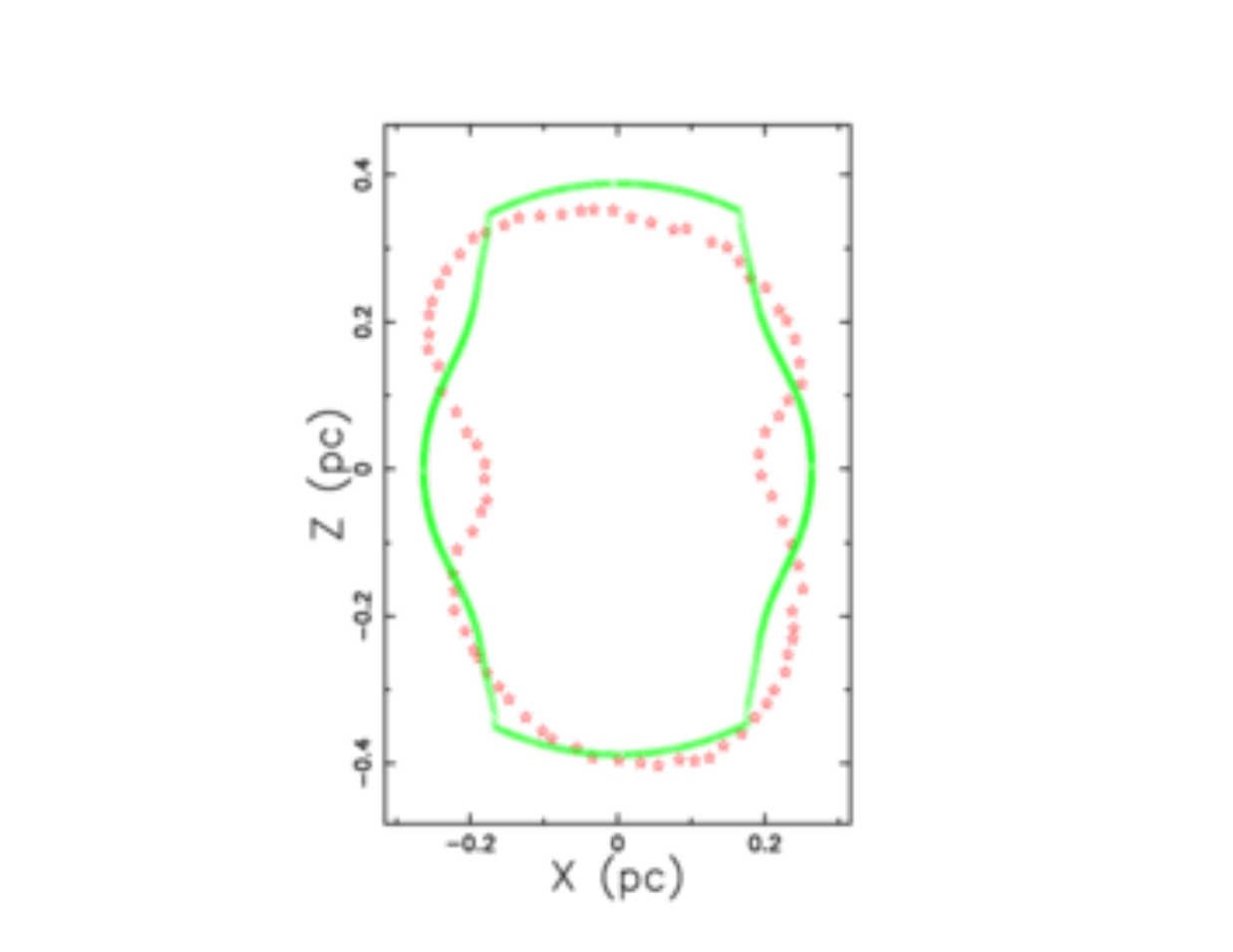}
\end {center}
\caption
{
Geometric section of \sn1987a
in the $x-z$ plane for  a toroidal    profile
(green points)
and the observed profile (red stars).
The parameters
$r_0=0.15 $ pc,
$R_T=1.02\,r_0$ pc,
$r_T=0.9\,R_T$ pc,
$\rho_c=1$,
$\rho_1=10 \,\rho_c$,
$t=27.7$  yr,
$t_0 =5.65$ yr,
$v_0=26000$ km s$^{-1}$
and 
$h=0.01$ yr  
give
$\epsilon_{\mathrm {obs}}=86.28\%$.
}
\label{cut_torus_1987a}
    \end{figure*}

\begin{figure*}
\begin{center}
\includegraphics[width=7cm,angle=-90]{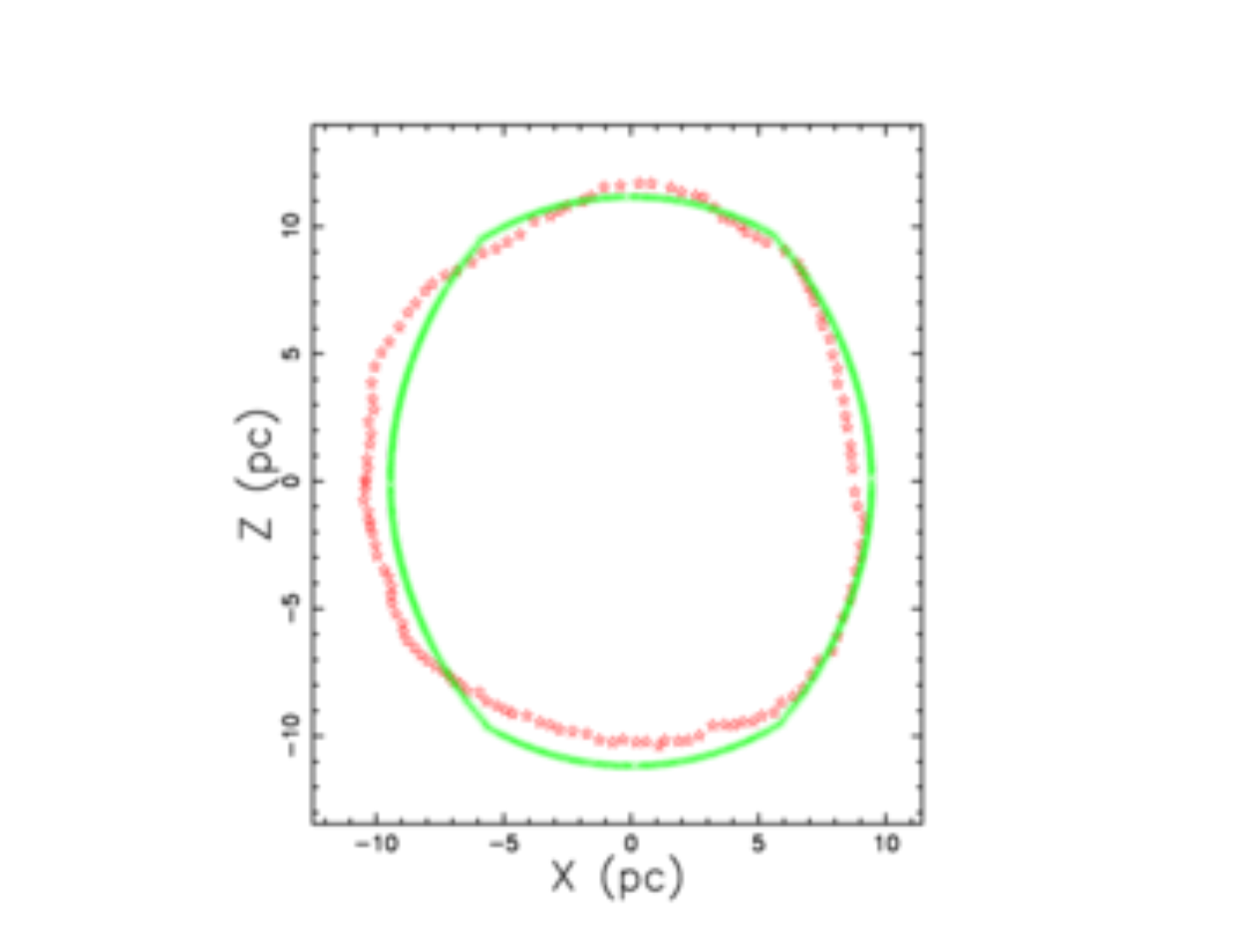}
\end {center}
\caption
{
Geometric section of \s1006
in the $x-z$ plane for  a toroidal  profile
(green points)
and the observed profile (red stars).
The parameters
$r_0=3.6  $ pc,
$R_T=1.02\,r_0$ pc,
$r_T=0.86\,R_T$ pc,
$\rho_c=1$,
$\rho_1=3 \,\rho_c$,
$t=1000$  yr,
$t_0 =135.64$ yr,
$v_0=26000$ km s$^{-1}$
and 
$h=1$ yr  
give
$\epsilon_{\mathrm {obs}}=94.41\%$.
}
\label{sn1006_vera_obs_torus}
    \end{figure*}

In the case of a toroidal   profile of
the density, we obtained a numerical  
solution  
for the two recursive equations (\ref{recursive1}) 
and (\ref{velocityeuler}).
Figure \ref{cut_torus_1987a} presents the results 
for    \sn1987a in the $x-z$ plane
and  Figure \ref{sn1006_vera_obs_torus} for \s1006.

% figure   cut_torus_1987a
% figure   cut_torus_1987a

% figure   sn1006_vera_obs_torus
% figure   sn1006_vera_obs_torus
A toroidal region  around the point of the explosion 
with density greater than the surrounding
medium produces theoretical sections 
which are comparable  
with  the observed sections 
of the two SNRS here simulated.

\section{Theory of the image}
\label{sec_energy_image}

The astronomical images are given by cut or   2D maps 
for  the intensity of emission.
The shape of the intensity of emission 
is a combination of different
processes which are as follows.
\begin{enumerate}
\item An assumption on the the transfer equation:
we  adopt an optically thin medium.   
\item Type of emission: we choose non thermal emission.  
\item Geometry  of the radiative zone:
      the zone of emission resides in the thin advancing 
      shell, which in our case is not spherical. 
\end{enumerate}
More details on the above assumptions can be found 
in Section 7 of \cite{Zaninetti2018a}.

\subsection{How to build an image}

\begin{figure*}
\begin{center}
\includegraphics[width=7cm]{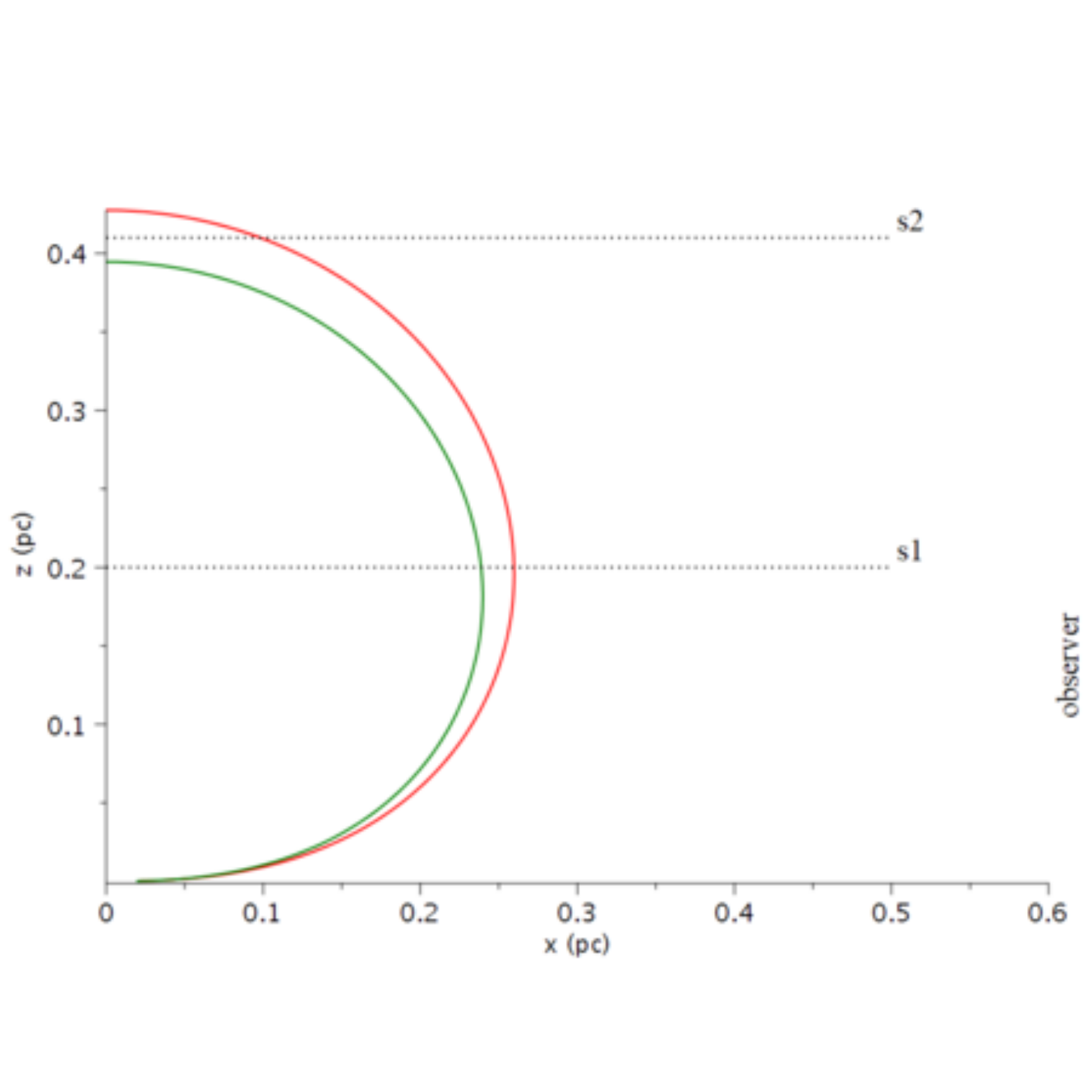}
\end {center}
\caption
{
First quadrant for the
layer between the upper (red curve) and the lower (green curve)  
section of \sn1987a
in the $x-z$ plane with an inverse square  profile
when    $0 < \theta < \pi$. 
Parameters as in Figure \ref{cut_invsquare_1987a}.
The observer is at infinity of the $x$-axis and two 
lines of sight, $s1$ and $s2$, are marked.
}
\label{observer_invsquare_sn1987a}
    \end{figure*}

The {\it first} model assumes that the intensity of the
image  is proportional to the length along the line of sight within the emitting region.
The derivation  of this length 
can be complicated, due to the complex morphology of the analysed 
object  
and to the infinite points of view of the observer.
Figure \ref{observer_invsquare_sn1987a} presents 
a sketch of the emitting layer, some lines of sight, 
and the position of the observer.
% figure   observer_invsquare_sn1987a
% figure  observer_invsquare_sn1987a
This sketch  allows building a geometric model
for the theoretical
image formed between the two layers.
The {\it second} model
for the intensity of the observed radiation  
assumes a     synchrotron source 
for luminosity proportional to the 
flux of kinetic energy, $L_m$,
\begin{equation}
L_m = \frac{1}{2}\rho A  V^3
\quad,
\label{fluxkineticenergy}
\end{equation}
where $A$ is the considered area, $V$ the velocity 
and $\rho$ the density, 
see formula (A28)
in \cite{deyoung}.
In our  case $A=R^2 \Delta \Omega$,
where $\Delta \Omega$ is the considered solid angle along 
the chosen direction.
This means
\begin{equation}
L_m = \frac{1}{2}\rho \Delta \Omega  R^2 V^3
\quad,
\label{fluxkinetic}
\end{equation}
where $R$  is the instantaneous radius of the SNR and
$\rho$  is the density in the advancing layer
in which the synchrotron emission takes place.
The   observed luminosity along a given direction 
can  be expressed as
\begin{equation}
L  = \epsilon_m  L_{m}
\label{luminosity}
\quad,
\end{equation}
where  $\epsilon_m$  is  a constant  of conversion
from  the mechanical luminosity   to  the
observed luminosity in the considered band.
The numerical algorithm which allows us to
build  a complex  image is  now 
outlined.
\begin{itemize}
\item  An empty (value=0)
memory grid  ${\mathcal {M}} (i,j,k)$ which  contains
$400^3$ pixels is considered.
\item  We  first  generate an
internal 3D surface by rotating the ideal image
of  $180^{\circ}$
around the polar direction and a second  external  surface at a
fixed distance $\Delta R$ from the first surface. As an example,
we fixed $\Delta R = R/12 $, where $R$ is the
momentary  radius of expansion.
The points on
the memory grid which lie between the internal and external
surfaces are stored in memory on
${\mathcal {M}} (i,j,k)$ with a variable integer
number   according to formula
(\ref{fluxkinetic}).
\item  Each point of
${\mathcal {M}} (i,j,k)$  has spatial coordinates $x,y,z$ which  can be
represented by the following $1 \times 3$  matrix, $A$,
\begin{equation}
A=
 \left[ \begin {array}{c} x \\\noalign{\medskip}y\\\noalign{\medskip}{
   z}\end {array} \right]
\quad.
\end{equation}
The orientation  of the object is characterized by
 the
Euler angles $(\Phi, \Theta, \Psi)$
and  therefore  by a total
$3 \times 3$  rotation matrix,
$E$, see \cite{Goldstein2002}.
The matrix point  is
represented by the following $1 \times 3$  matrix, $B$,
\begin{equation}
B = E \cdot A
\quad.
\end{equation}
\item 
The intensity map is obtained by summing the points of the
rotated images
along a particular direction.
\item 
The effect of the  insertion of a threshold intensity, $I_{tr}$,
given by the observational techniques,
is now analysed.
In both models
the threshold intensity can be
parametrized  by  $I_{max}$,
the maximum  value  of intensity
characterizing the map.
\end{itemize}
More details on the theory of the image 
can be found in \cite{Zaninetti2018a}.

\subsection{Results for the first model}

\begin{figure*}
\begin{center}
\includegraphics[width=7cm,angle=-90]{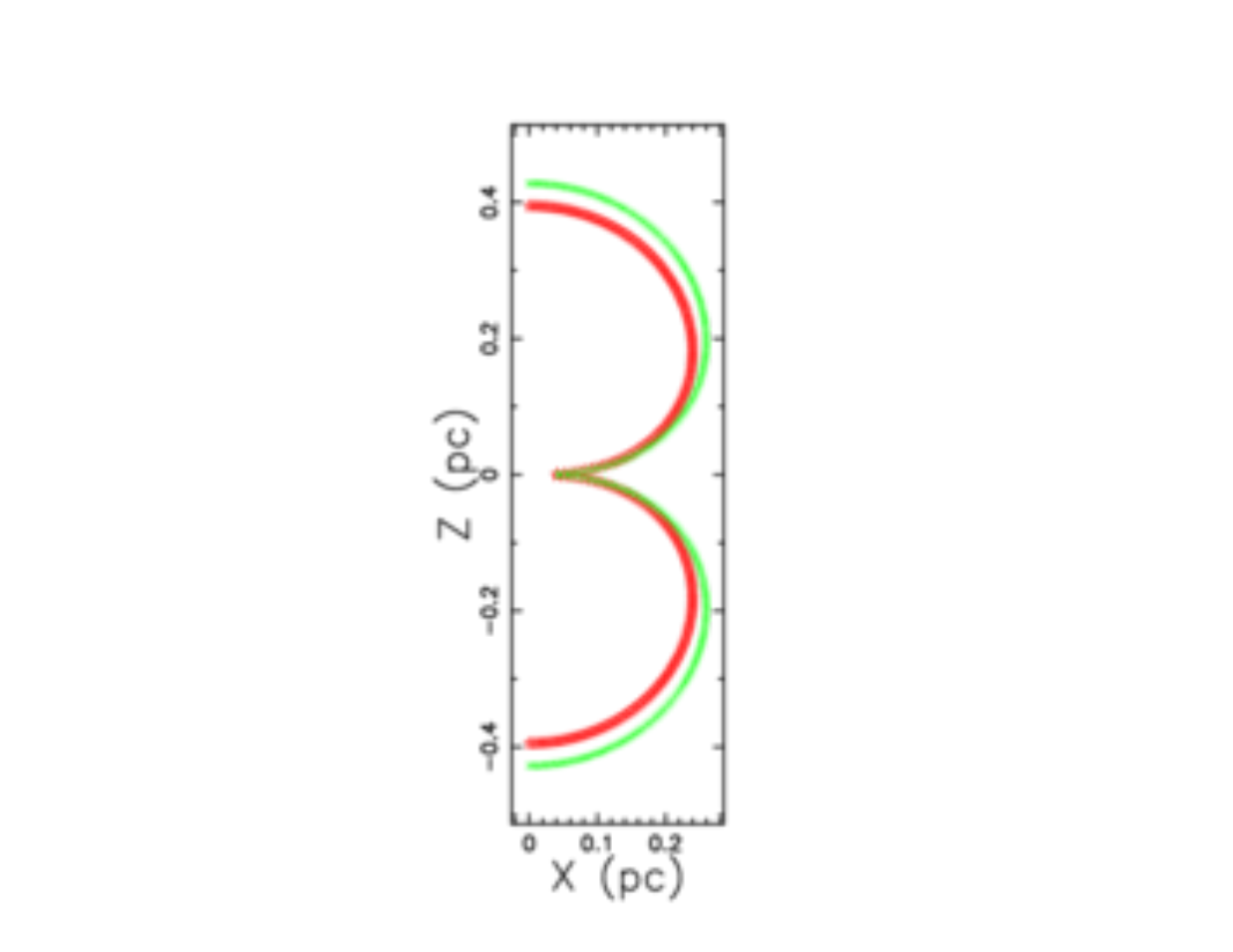}
\end {center}
\caption
{
The layer between the upper (green points) and the 
lower (red points)  
section of \sn1987a
in the $x-z$ plane with an inverse square  profile
when   $0 <\theta < \pi$. 
Parameters as in Figure \ref{cut_invsquare_1987a}.
}
\label{strato_sn1987a_invsquare}
    \end{figure*}

\begin{figure*}
\begin{center}
\includegraphics[width=7cm,angle=-90]{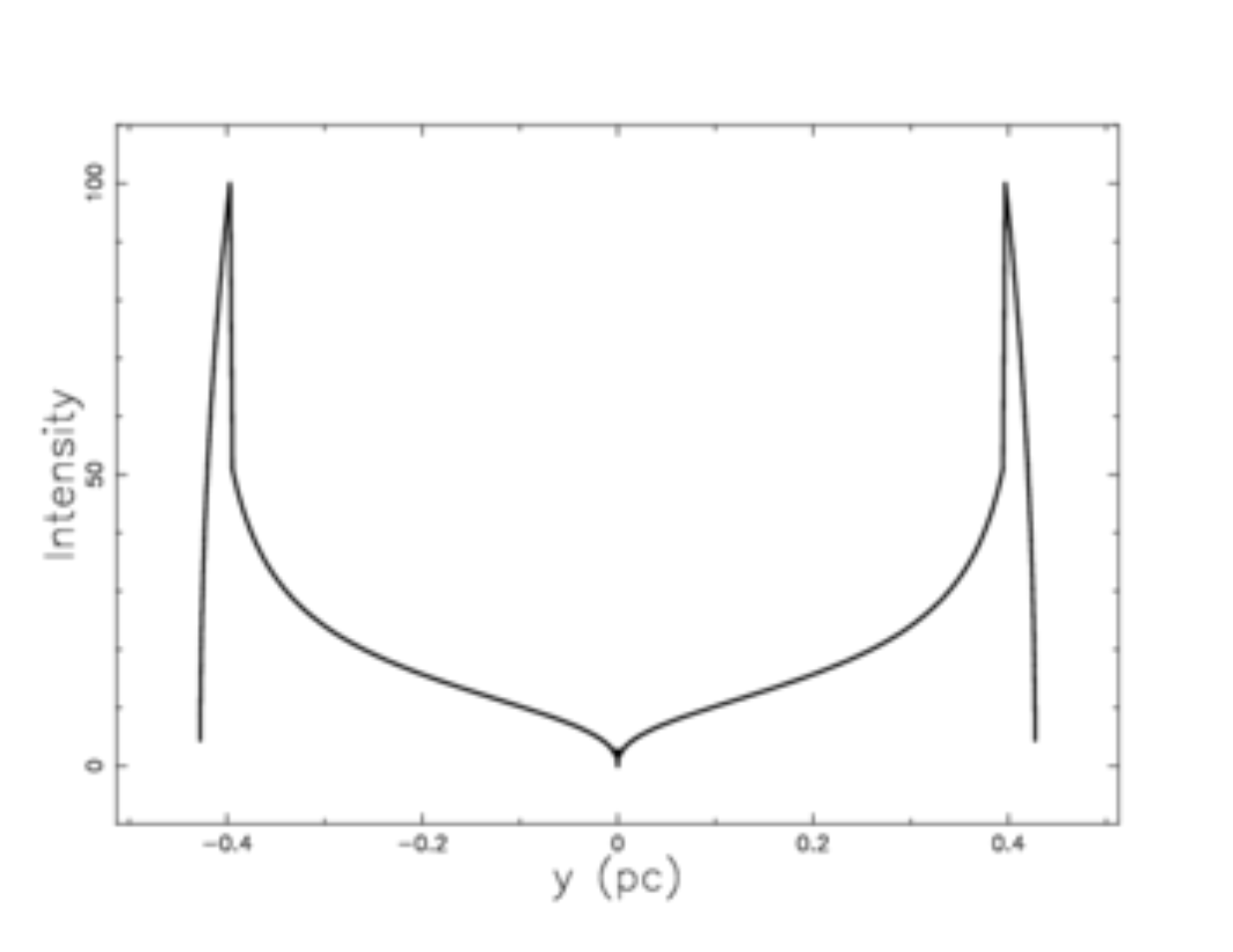}
\end {center}
\caption
{
Cut of the mathematical  intensity ${I}$,
as a function of the polar  $z$-axis for  \sn1987a with an inverse square  
profile.
Parameters as in Figure \ref{cut_invsquare_1987a} and $N=240$.
}
\label{theoreticalcut_sn1987a_invsquare}
    \end{figure*}

The radiation of the SN is supposed  to originate 
from between a lower boundary of radius  $r_{inf}$, 
given as an example  
by $r(t;z0,r_0,\theta)$ of equation (\ref{rtinversesquare}),
and an upper boundary with radius given by the equation
\begin{equation}
r_{sup} = r_{inf} + \frac{r_{inf}}{12}
\quad.
\end{equation}  
The above layer is visualized 
in Figure \ref{strato_sn1987a_invsquare} 
for \sn1987a when an inverse square profile is adopted.
% figure   strato_sn1987a_invsquare
% figure   strato_sn1987a_invsquare
%siamoqui
The image of \sn1987a is formed between the two layers.
The data of the two layers  are  stored in memory  
as Cartesian coordinates
\begin{subequations}
\begin{align}
x_{sup}(n) = r_{sup}(n) \sin (\theta(n))
\\
z_{sup}(n) = r_{sup}(n) \cos (\theta(n))
\\
x_{inf}(n) = r_{inf}(n) \sin (\theta(n))
\\
z_{inf}(n) = r_{inf}(n) \cos (\theta(n))
\end{align}
\end{subequations}
where $n$ is the considered index.
In order to build the {\it first} model 
we assume   that  $z$ varies between 
$z_{sup}(1)$  and $z_{sup}(N)$  
where $N$ represents the number of points.
The image is assumed to be proportional to the 
distance between
the $x$ coordinate on  the  upper layer 
and the $x$ coordinate on the lower layer as a function of the
position $z$ on the polar axis, see Figure
\ref{theoreticalcut_sn1987a_invsquare}.
% figure   theoreticalcut_sn1987a_invsquare
% figure   theoreticalcut_sn1987a_invsquare
In principle, astronomers should produce
these observational cuts in order to compare the 
observations with the theory here developed.

\begin{figure*}
\begin{center}
\includegraphics[width=7cm,angle=-90]{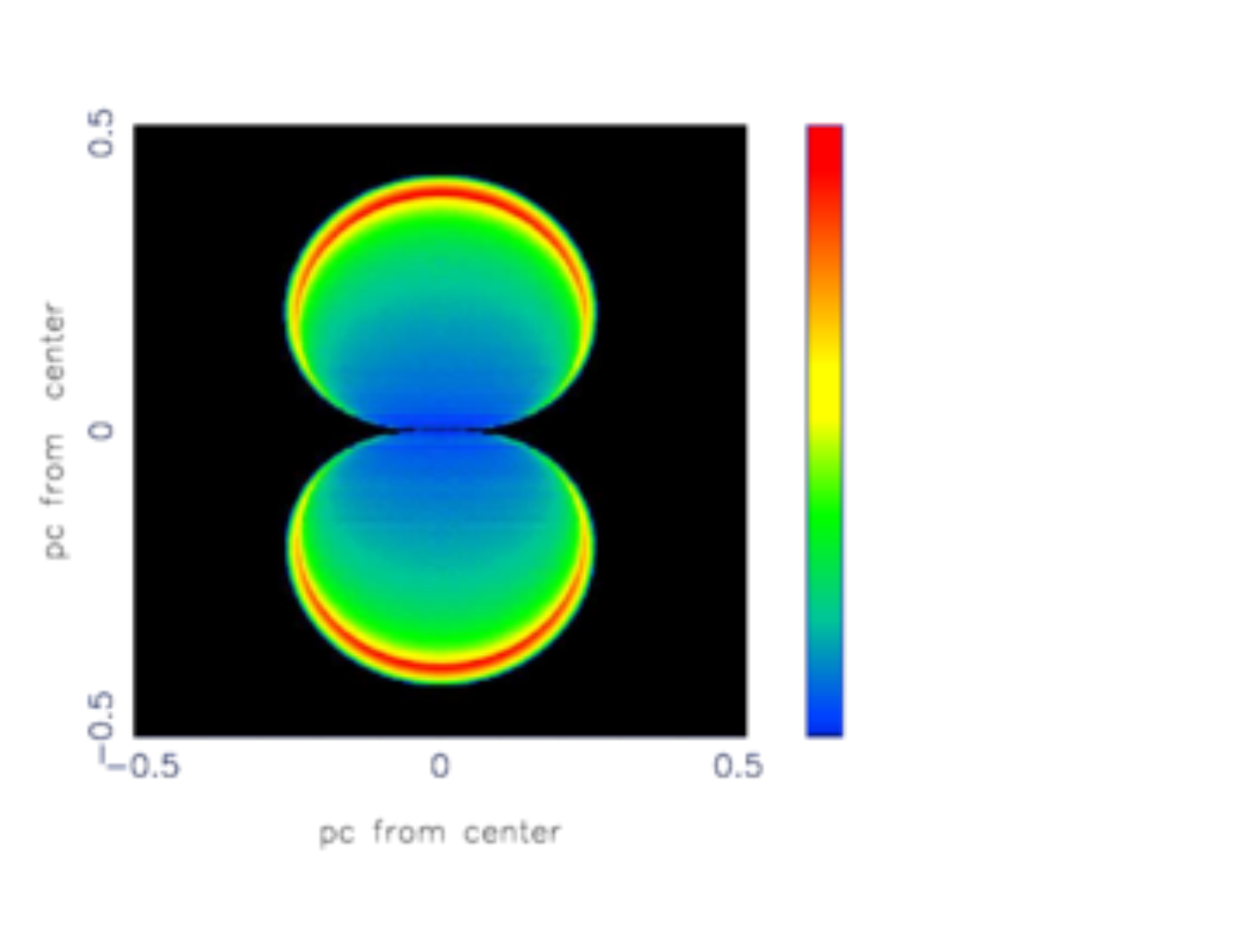}
\end {center}
\caption
{
Map of the theoretical  intensity ${I}$,
for  \sn1987a with an inverse square  
profile made by 400$\times$400 pixels. 
Parameters as in Figure \ref{cut_invsquare_1987a}.
}
\label{invsquare_1987a_heat_linea_yz}
    \end{figure*}

The same algorithm allows building  a map 
of the theoretical intensity for the rotated image of \sn1987a,
see Figure \ref{invsquare_1987a_heat_linea_yz}.
% figure   invsquare_1987a_heat_linea_yz
% figure   invsquare_1987a_heat_linea_yz
The observed image of \sn1987a 
should be digitized 
in order to make a comparison with the theory.

\subsection{Results for the second model}

\begin{figure}
  %\begin{center}
\includegraphics[width=6cm,angle=-90]{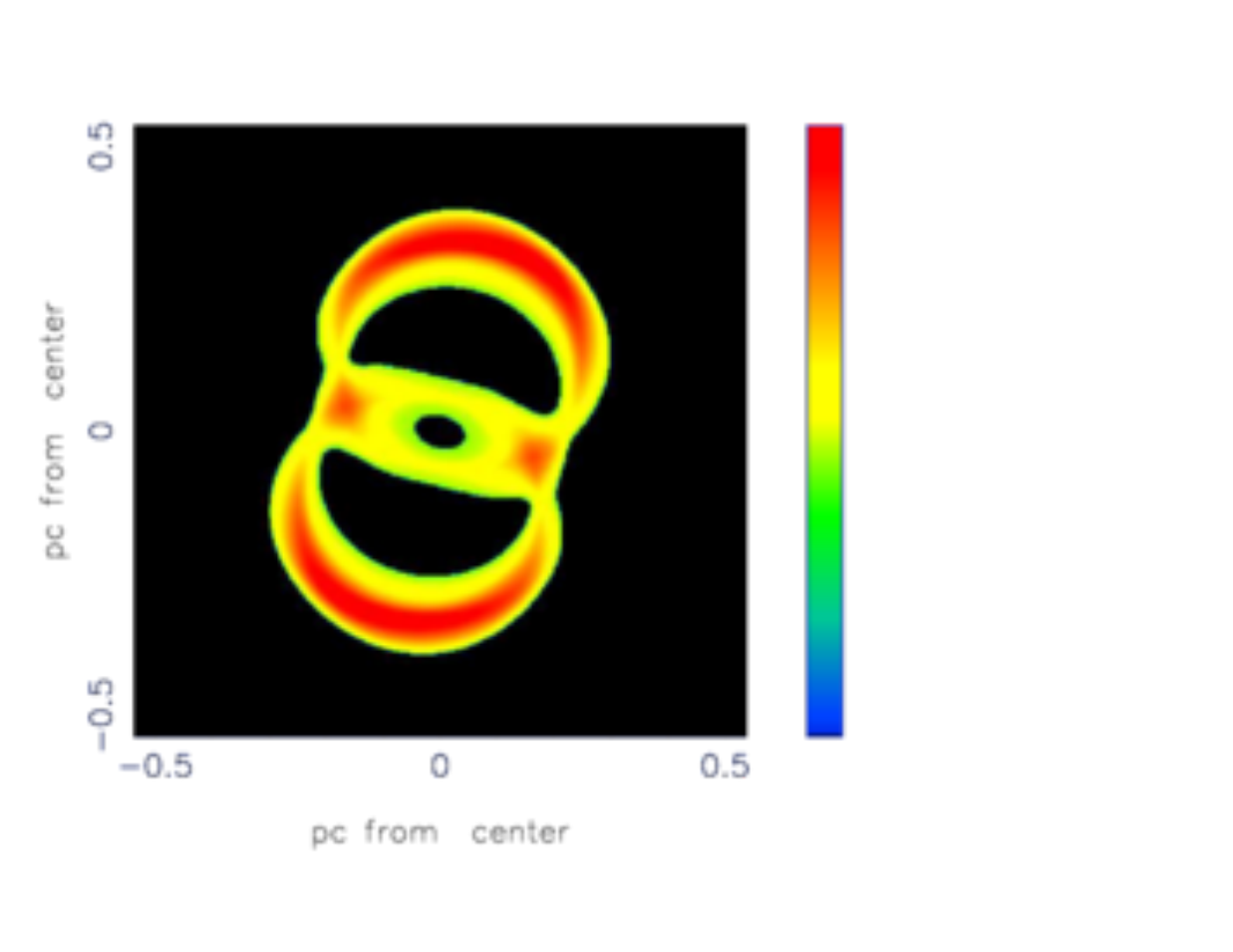}
  %\end{center}
\caption {
Model map of \sn1987a rotated in
accordance with the observations,
for an inverse square  medium with
parameters as in Figure \ref{cut_invsquare_1987a}.
The three Euler angles
characterizing the orientation
are
     $ \Phi   $= 105 $^{\circ }$,
     $ \Theta $=  45 $^{\circ }$
and  $ \Psi   $=-165 $^{\circ }$.
In this map $I_{tr}= I_{max}/2.2$.
}
    \label{invsquare_sn1987a_hole}
    \end{figure}

\begin{figure}
  %\begin{center}
\includegraphics[width=6cm,angle=-90]{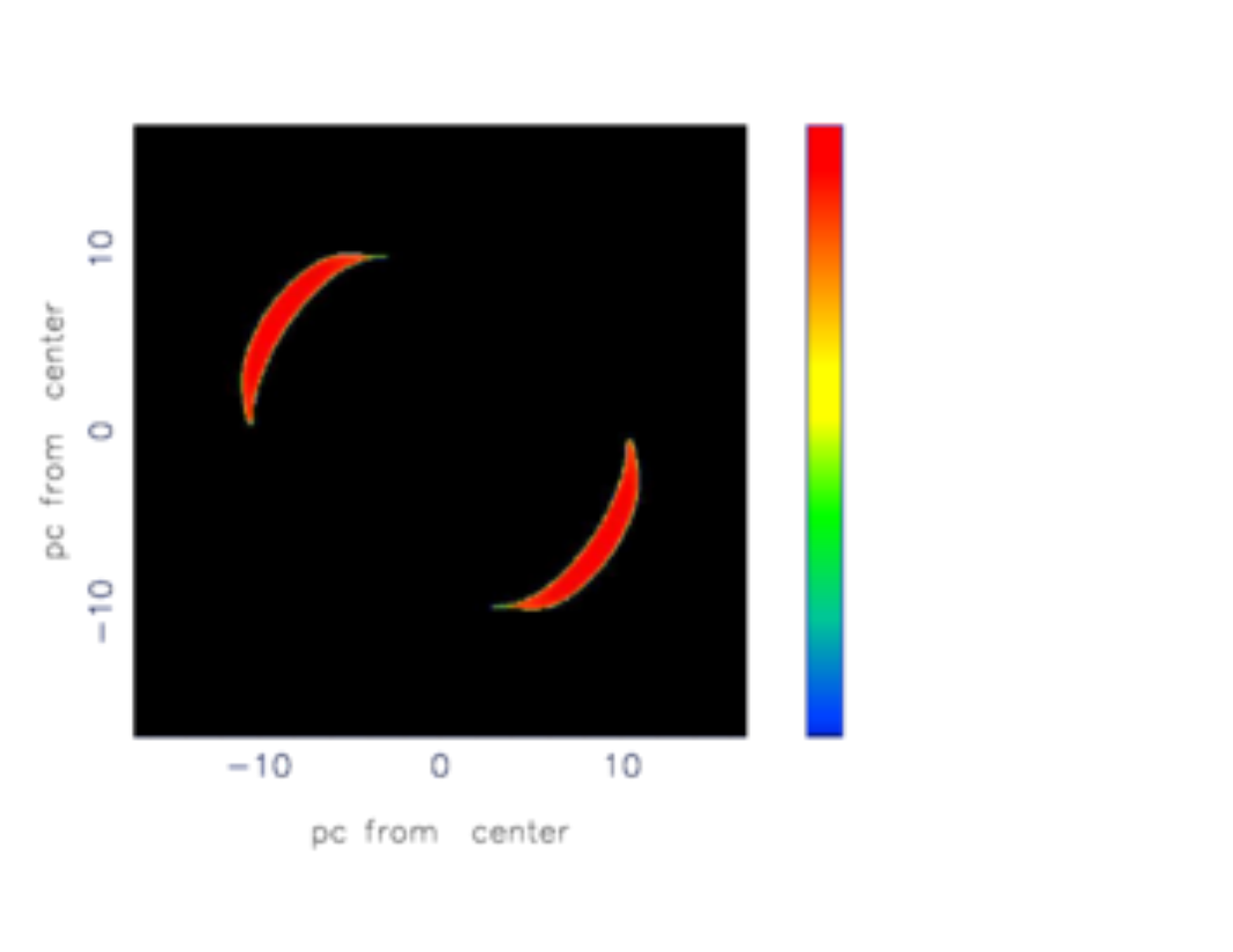}
  %\end{center}
\caption {
Model map of \s1006 rotated in
accordance with the $\gamma$-ray observations with HESS,
for a toroidal medium.
Physical parameters as in Figure \ref{sn1006_vera_obs_torus}.
The three Euler  angles characterizing
the orientation of the observer
are
     $ \Phi   $=180  $^{\circ }$,
     $ \Theta $=90   $^{\circ }$
and  $ \Psi   $=-55  $^{\circ }$.
In this map $I_{tr}= I_{max}/1.2$.
          }%
    \label{torus_sn1006_hole_rotated}
    \end{figure}

This  second model allows simulating
particular effects, such as
the triple ring system of \sn1987a, see
Figure  \ref{invsquare_sn1987a_hole},
where three zones or holes  with theoretical intensity
under the threshold value  are visible.
% figure  invsquare_sn1987a_hole
% end figure invsquare_sn1987a_hole
%siamoqui
The enigma of the three holes is solved.
Characteristic features,  such as  the `jet  appearance' visible in some
maps for \s1006, see the X-ray map  \ref{SN_CHANDRA_X_2013}  
and $\gamma$-ray map \ref{SN1006_HESS},
are theoretically modeled  in Figure \ref{torus_sn1006_hole_rotated}.
% figure  torus_sn1006_hole_rotated
% end figure torus_sn1006_hole_rotated
The enigma of the jet, which is a cone,
in a nearly spherical expansion is 
solved.

\section{Conclusions}

{\bf Models}
We have adopted four models in the framework 
of the conservation of energy for  
the thin layer approximation
and applied them to
two SNRs; the results for the percentage reliability,
see formula (\ref{efficiencymany})
are shown in Table \ref{synoptic}.
%inizio table synoptic 
\begin{table}[ht!]
\caption {
Synoptic  percentage reliability   for  the best model
of  SNRs with different 
profiles for the density.
}
\label{synoptic}
\begin{center}
\begin{tabular}{|c|c|c|}
\hline
Model  & SNR       & $\epsilon_{\mathrm {obs}}(\%)$ \\
\hline                                                   
inverse~square~profile &  {SN \,1987A\,}  & 91.92  \\ 
\cline{2-3}
~             &  {SN \,1006\,}   & 94.34  \\
\hline
power~profile &  {SN \,1987A\,}  & 91.92  \\  
\cline{2-3}
~ &  {SN \,1006\,}   & 93.56  \\
\hline
exponential~profile &  {SN \,1987A\,}  & 92.45  \\  
\cline{2-3}
~ &  {SN \,1006\,}   & 94.26  \\
\hline
toroidal~profile &  {SN \,1987A\,}  & 86.28  \\  
\cline{2-3}
~ &  {SN \,1006\,}   & 94.41  \\
\hline
\end{tabular}
\end{center}
\end{table}
The best fit for  the asymmetric section of   \sn1987a 
is represented  by the exponential model  for  decreasing 
density in the polar direction
and     by the toroidal profile of the density for  \s1006.

{\bf Image theory}
Two models for the intensity of the image in an SNR  have been presented, 
which both require an  analytical  or numerical 
function for  the advancing radius in terms of the polar angle.
The appearance of the triple ring in \sn1987a, 
see Figure \ref{invsquare_sn1987a_hole}, and
the jet-feature in \s1006, 
see Figure \ref{torus_sn1006_hole_rotated},
have been simulated.

{\bf Back reaction}

The derivation of an approximate form for the 
losses, see equation (\ref{losses}),
allows deriving a finite rather than infinite
length for an  SNR, see equation (\ref{finitelength}).

\section*{Acknowledgments}

At the time of writing, an animated version of 
Figure \ref{3dinversesquaresn1987a} is 
visible at
\\
\url{http://personalpages.to.infn.it/~zaninett/image/sn1987a_animation.gif}.

%\bibliography{biblio}
\providecommand{\newblock}{}

\end{document}